%% file: Long-run_patterns_in_the_discovery_of_the_adjacent_possible_for_arxiv_2023.tex
\documentclass[english,a4paper, 12pt,
twocolumn]{article}
\usepackage[hmargin={0.6in,0.6in},vmargin={1in,1in}]{geometry}
\usepackage[OT1]{fontenc}
\usepackage[latin9]{inputenc}
\usepackage{pbox}
\usepackage{array}
\usepackage{rotating}
\usepackage{multirow}
\usepackage{makecell}
\usepackage{amsbsy}
\usepackage{amsmath}
\usepackage{authblk}
\usepackage{stackengine, graphicx}
\usepackage{setspace} 

\usepackage{natbib}
\usepackage{caption}
\usepackage{subcaption}
\usepackage[switch,columnwise]{lineno}
\usepackage{hyperref}
\usepackage{threeparttable}
\usepackage{fancyhdr}
\usepackage{babel}
\makeatletter
\usepackage{array} 
\newcolumntype{H}{>{\setbox0=\hbox\bgroup}c<{\egroup}@{}}

\usepackage{tikz}
\usetikzlibrary{calc}
\usetikzlibrary{decorations.pathmorphing}
\usepackage{lettrine}

\makeatother
\date{
}

\author[1]{Josef Taalbi}
    
\affil[1]{\footnotesize{Department of Economic History, Lund University. Corresponding author: josef.taalbi@ekh.lu.se}}

\begin{document}

\twocolumn[
 \begin{@twocolumnfalse}
\title{
Long-run patterns in the discovery of the adjacent possible}
\maketitle

\textbf{
\linebreak
The notion of the ``adjacent possible'' has been advanced to theorize the generation of novelty across many different research domains. This study is an attempt to examine in what way the notion can be made empirically useful for innovation studies. A theoretical framework is construed based on the notion of innovation a search process of recombining knowledge to discover the ``adjacent possible''. The framework makes testable predictions about the rate of innovation, the distribution of innovations across organizations, and the rate of diversification or product portfolios. The empirical section examines how well this framework predicts long-run patterns of new product introductions in Sweden, 1908-2016 and examines the long-run evolution of the product space of Swedish organizations. The results suggest that, remarkably, the rate of innovation depends linearly on cumulative innovations, which explains advantages of incumbent firms, but excludes the emergence of ``winner takes all'' distributions. The results also suggest that the rate of development of new types of products follows ``Heaps' law'', where the share of new product types within organizations declines over time. The topology of the Swedish product space carries information about future product diversificaitons, suggesting that the adjacent possible is not altogether ``unprestatable''.} 

 \end{@twocolumnfalse}
\vspace{1cm}
]

\section{INTRODUCTION}
\doublespacing
It has long been recognized that the emergence of novelty is a core aspect of evolutionary processes, propelling language, ecological, social, technological and economic systems towards new constellations. A branch of studies have attempted to theorize the emergence of novelty across these different fields and, with some degree of success, also investigate the presence of common general patterns and mechanisms. 

Up until recently, most cross-field talk between evolutionary biology and innovation studies or evolutionary economics \citep{klepper1997, nelson1982} has been centered on variety and selection, and evolution as adaptive hill climbing on fitness landscapes \citep{kauffman1993, levinthal1997}, or evolution as a form of ``tinkering'' or trial and error process \citep{jacob1977, wagner2014}. While broadly useful, the focus on adaptive search is however restrictive if the goal is to explain the emergence of novelty \citep{felin2014}. In the economics of innovation, it has long been recognized that innovation are new (re-)combinations of previous innovations, that come into economic use (\citealt{schumpeter1911}; compare \citealt{oecd2005}). In keeping with this, the innovation process is usually conceptualized as a process of search to find better combinations \citep{weitzman1998, fleming2001, arthur2007, arthur2009}. Furthering such intuitions, the concept of the ``adjacent possible'' was introduced by Stuart \citet{kauffman2000} to explain the emergence of novelty in complex adaptive systems such as the biosphere, where life creates new niches and opportunities into which it expands. The notion of an adjacent possible divides the space of innovations into three conceptual categories: 
\begin{itemize}
\item[i]	those that have been discovered, 
\item[ii]	those that can currently be discovered from (recombining) those that have already been discovered
\item[iii]	innovations that are, as it were, ``out of reach'', but may become possible to discover in the future.
\end{itemize}

This framework is simple but powerful. Recently, a number of studies have drawn on this notion to examine how novelties give rise to other novelties, suggesting the existence of statistical laws for the rate of novelty generation in a broad set of phenomena \citep{loreto2016, tria2014, iacopini2018, ubaldi2021}. Specifically, the so-called urn models of \citet{tria2014} and \citet{loreto2016} model the generation of novelty by assuming that novelty can be represented by balls with different colors in an urn, showing how novelties (new colors) can trigger other novelties through a simple reinforcement mechanism, viz. that a drawn ball with a certain color will increase the probability that it is drawn again. These models predict statistical laws for the rate at which novelties happen (known as Heaps' law, see Heaps 1978) and the frequency distribution of different types of novelties (colors drawn), known as Zipf's law. Other models have explored novelty generation on social and innovation networks \citep{iacopini2018, ubaldi2021}. These studies showcase a pathway to better understand patterns in the emergence of novelties across a broad set of systems. 

A challenge is how to operatinalize the adjacent possible, seeing as it is fundamentally ``unprestatable'' \citep{kauffman2019}, meaning that it is impossible to predict what novelties will crop up. And more troubling: there is no way of directly estimating the size of the adjacent possible at one point in time, even in hindsight, unless all of the possibilities in the adjacent possible were in fact discovered. Recent studies have argued that one may still in principle model the adjacent possible on the basis of the number of recombinations that are possible to make from present, available, knowledge \citep{steel2020, koppl2021, cortes2022}. This model, called the Theory of the Adjacent Possible (TAP), strongly suggests super-exponential rates of innovation in the long run, which for example aligns the with the ``hockey-stick'' shape of long-run GDP growth \citep{steel2020, koppl2021}.

Even so, there are, however, two major challenges of how to operationalize the concept of the adjacent possible. 
Firstly, as noted recently, the urn model \citep{tria2014, loreto2016} is restricted as there is no ``cross-talk'' between product types, or ``elements''  \citep[p.~139]{kauffman2019}.\footnote{The full quote: ``The model is lovely, but does not yet answer our needs, for it is one of a branching set of independent lineages of descendant colored balls. A red ball gives rise to an orange ball, which gives rise to a blue ball. There is no cross-talk between lineages augmenting the combinatorial formation of new colors as there is in the economic web's evolution with new complements and substitutes arising from old ones by new jury-rigged combinations of one or several prior goods. I hope that a good model or set of models can be constructed'' \citep[p.~139]{kauffman2019}.} There is in other words a challenge in reuniting the predictions of \citet{tria2014}, with several other observations from theory, namely that innovation is the result of recombinant search. While the TAP framework is based on recombinant growth it has not been shown to reproduce Heaps' law, or other statistical regularities as regards the emergence of novelty.

Secondly, in the present context of long-run innovation dynamics, the recombinant perspective leads to a puzzle. If the adjacent possible of organizations grows exponentially (or, indeed, super-exponentially, compare \citealt{sole2016,steel2020, cortes2022}) with its innovation experience, and organizations innovate at a rate governed by the adjacent possible, it is easy to see that the long-run industrial dynamics should give way to a winner-takes-all phenomenon (for a formal proof of the link between super-linear attachment kernels and winner-takes-all distributions, see \citealt{krapivsky2001}). This does not align with various studies that have argued that long-run innovation is subject to natural/physical constraints \citep{ayres1994}, resource constraints \citep{weitzman1998}, or towering complexity \citep{strumsky2010, arnold2019, bloom2020}.
This suggests that the TAP framework makes too strong predictions for some applications, and that the probing of the adjacent possible has considerable constraints, as suggested by \citet{weitzman1998}, for example in terms of limited capability, resources and search routines available to agents such as organizations. 

This study takes up the challenge of how to theorize the growth of the adjacent possible in a more flexible way. The aim of this study is twofold. This study aims to accomplish a unification of the concepts of recombinant search, the limitations introduced by \citet{weitzman1998} and the notion of the adjacent possible, to make testable predictions about the long-run rate of innovation, the distribution across firms and the rate of diversifications of firms.

Secondly, I examine whether this framework is useful in explaining empirical patterns of innovation, by leveraging historical data on new product introductions in the Swedish engineering industry during the period of 1908-2016 \citep{kander2019,taalbi2021}  in long-run innovation, including the rate of innovation, distribution across organizations, and patterns of diversification. In addition, I also look at the product space of co-development of innovations to discuss the structure of the adjacent possible, which may be inform further research on this topic. 

The rest of the study is organized as follows. In section \ref{sec:theory}, a theoretical framework is proposed that unites the notion of the adjacent possible, with innovation as recombination \citep{kauffman1993, weitzman1998, arthur2009}, in its weak or strong form. The model produces testable predictions as regards the rate of innovation over time and across organizations and the diversification of products, as also suggested by \citep{tria2014}. Section \ref{sec:data} introduces the data on Swedish innovation output and section \ref{sec:results} analyzes the rate of innovation, product diversification and product network for Swedish organizations 1908-2016. Section \ref{sec:discussion} concludes.

\section{A THEORETICAL FRAMEWORK}
\label{sec:theory}

\subsection{Main framework}
\label{sec:main}

To build a framework I depart from the following considerations. The core assumption is that innovations embody different types of knowledge. Each organization has a number of product types in their repertoire. From the organization's point of view, new product types expand the organization's knowledge base to new fields \citep{katila2002, march1991}. 
Each organization also has a history of product improvements made in the respective fields, representing advances in knowledge in those fields. The number of product types in the organization's repertoire one may call its product diversity $D$. The number of improvements is $I$. The total cumulated number of innovations is $k=D+I$, embodying the cumulated knowledge base of the organization.

This framework should be able to deliver predictions about three aspects of long-run innovation dynamics: 

\begin{itemize}
\item	The contribution of cumulative innovations to the organization's rate of innovation $dk/dt$
\item	The rate of diversification of the product portfolio ($dD/dk$)
\item	The relative frequency of innovations across organizations, $P(k)$
\end{itemize}

To make these predictions, this study makes use of the notion of the adjacent possible as discussed previously. 
For each organization there is a set of product types ${\cal{U}}_D$, not yet produced but possible to produce given the current knowledge base. There is also a set of product improvements ${\cal{U}}_I$, not yet developed, that an organization can develop given its current knowledge base. ${\cal{U}}={\cal{U}}_I \cup {\cal{U}}_D$ is the set of all adjacent possible innovations. 

The question is now only how, more precisely, the set of adjacent possible innovations depends on the set of available knowledge of the organization, embodied in earlier innovations. The key is \emph{recombination}.
 To fix ideas, Figure 1 illustrates the interaction of the set of drawn elements ${\cal{S}}$ and the adjacent possible innovations ${\cal{U}}$ in an urn model where agents search by recombining from a subspace of drawn elements. 

\begin{figure*}
\centering
\begin{tikzpicture}
\node[] at (0,2) {$t$};
\node[] at (4+2,2) {$t+1$};
\node[] at (-1,1) {$\mathcal{S}$};
\node[] at (3+2,1) {$\mathcal{S}$};
\node[] at (-1,-3) {$\mathcal{U}$};
\node[] at (3+2,-3) {$\mathcal{U}$};
\node[] at (2.3,2.0) {$\mathcal{R}$};

\node [draw,
	rectangle, rounded corners=10,
	minimum size =3cm,
	label={}] (A) at (0,0){};

\node [draw,
	rectangle,
	rounded corners=10,
	dashed,
	fill = blue!10!white,
	minimum width =1.4cm,
	minimum height = 2*0.3cm,
	label={}] (Ax) at (0.3+0.3,-0.9){} ;

\node [draw,
	rectangle,
	rounded corners=10,
	dashed,
	fill = blue!10!white,
	minimum width =1.4cm,
	minimum height = 2*0.3cm,
	label={}] (Rx) at (0.3+0.3+2,-1.1){} ;

\node [draw,
	rectangle, 
	rounded corners=10,
	minimum size =1.7cm,
	label={}] (D_S) at (0.5,-0.5){};

\node [draw,
	circle, fill = green,
	minimum size =0.01cm,
	label={}] (A1) at (-0.7,0.7){};

\node [draw,
	circle, fill = violet,
	minimum size =0.01cm,
	label={}] (A2) at (0.4,0){};

\node [draw,
	circle, fill = yellow,
	minimum size =0.01cm,
	label={}] (A3) at (-1.0 ,-0.8){};

\node [draw,
	circle, fill = black,
	minimum size =0.01cm,
	label={}] (A4) at (0.9,-0.9){};

\node [draw,
	circle, fill = gray,
	minimum size =0.01cm,
	label={}] (Aa) at (0.2,1.0){};

\node [draw,
	circle, fill = blue,
	minimum size =0.01cm,
	label={}] (Ab) at (1.0,-0.11){};

\node [draw,
	circle, fill = orange,
	minimum size =0.01cm,
	label={}] (Ad) at (0.3,-0.9){};

\node [draw,
	circle, fill = white,
	minimum size =0.01cm,
	label={}] (Ae) at (-0.7,-1.2){};

\node [draw,
	rectangle, rounded corners=10,
	minimum size =3cm,
	label={}] (B) at (4+2,0){};

\node [draw,
	circle, fill = green,
	minimum size =0.01cm,
	label={}] (B1) at (-0.7+4+2,0.7){};

\node [draw,
	circle, fill = violet,
	minimum size =0.01cm,
	label={}] (B2) at (0.4+4+2,0){};

\node [draw,
	circle, fill = yellow,
	minimum size =0.01cm,
	label={}] (B3) at (-1.0+4+2,-0.8){};

\node [draw,
	circle, fill = black,
	minimum size =0.01cm,
	label={}] (B4) at (0.9+4+2,-0.9){};

\node [draw,
	circle, fill = gray,
	minimum size =0.01cm,
	label={}] (Ba) at (0.2+4+2,1.0){};

\node [draw,
	circle, fill = blue,
	minimum size =0.01cm,
	label={}] (Bb) at (1.0+4+2,-0.11){};

\node [draw,
	circle, fill = orange,
	minimum size =0.01cm,
	label={}] (Bd) at (0.3+4+2,-0.9){};

\node [draw,
	circle, fill = white,
	minimum size =0.01cm,
	label={}] (Be) at (-0.7+4+2,-1.2){};

\node [draw,
	circle, fill = red,
	minimum size =0.01cm,
	label={}] (Bf) at (-1+4+2.5,0){};


\node [draw,
	rectangle, rounded corners=10,
	minimum size =3cm,
	label={}] (C) at (0,-4){};

\node [draw,
	circle, fill = blue,
	minimum size =0.01cm,
	label={}] (C1) at (-0.3,-4+0.2){};

\node [draw,
	circle, fill = red,
	minimum size =0.01cm,
	label={}] (C2) at (0.3,-4+0.2){};

\node [draw,
	circle, fill = blue!10!white,
	minimum size =0.01cm,
	label={}] (C3) at (0.3,-4+-0.4){};

\node [draw,
	circle, fill = green!40!blue,
	minimum size =0.01cm,
	label={}] (C4) at (-0.3,-4-0.4){};

\node [draw,
	rectangle, rounded corners=10,
	minimum size =3cm,
	label={}] (D) at (4+2,-4){};

\node [draw,
	rectangle,
	rounded corners=10,
	dashed,
	minimum width =2cm,
	minimum height = 0.7cm,
	label={}] (Dx) at (-0.3+0.7+4+2,-4+0.7){} ;

\node [draw,
	circle, fill = blue,
	minimum size =0.01cm,
	label={}] (D1) at (-0.3+4+2,-4+0.2){};

\node [draw,
	circle, fill = red,
	minimum size =0.01cm,
	label={}] (D2) at (0.3+4+2,-4+0.2){};

\node [draw,
	circle, fill = blue!10!white,
	minimum size =0.01cm,
	label={}] (D3) at (0.3+4+2,-4+-0.4){};

\node [draw,
	circle, fill = blue!40!green,
	minimum size =0.01cm,
	label={}] (D4) at (-0.3+4+2,-4-0.4){};

\node [draw,
	circle, fill = red,
	minimum size =0.01cm,
	label={}] (Da) at (-0.3+4+2,-4+0.7){};

\node [draw,
	circle, fill = red,
	minimum size =0.01cm,
	label={}] (Db) at (0.3+4+2,-4+0.7){};

\node [draw,
	circle, fill = pink,
	minimum size =0.01cm,
	label={}] (Dc) at (0.3+4.8+2,-4+0.7){};

\node [draw,
	rectangle,
	rounded corners=10,
	dashed,
	minimum width =1.4cm,
	minimum height = 4*3*0.3cm,
	label={}] (R1) at (0.3+0.3+2,0.4){} ;

\node [draw,
	circle, fill = violet,
	minimum size =0.01cm,
	label={}] (R2) at (0.3+2,1.4){};

\node [draw,
	circle, fill = blue,
	minimum size =0.01cm,
	label={}] (R3) at (0.3*3+2,1.4){};

\node [draw,
	circle, fill = violet,
	minimum size =0.01cm,
	label={}] (R4) at (0.3+2,1.4-0.5){};

\node [draw,
	circle, fill = orange,
	minimum size =0.01cm,
	label={}] (R5) at (0.3*3+2,1.4-0.5){};

\node [draw,
	circle, fill = violet,
	minimum size =0.01cm,
	label={}] (R6) at (0.3+2,1.4-1.0){};

\node [draw,
	circle, fill = black,
	minimum size =0.01cm,
	label={}] (R7) at (0.3*3+2,1.4-1.0){};

\node [draw,
	circle, fill = blue,
	minimum size =0.01cm,
	label={}] (R8) at (0.3+2,1.4-1.5){};

\node [draw,
	circle, fill = orange,
	minimum size =0.01cm,
	label={}] (R9) at (0.3*3+2,1.4-1.5){};

\node [draw,
	circle, fill = blue,
	minimum size =0.01cm,
	label={}] (R10) at (0.3+2,1.4-2.0){};

\node [draw,
	circle, fill = black,
	minimum size =0.01cm,
	label={}] (R11) at (0.3*3+2,1.4-2.0){};

\node [draw,
	circle, fill = orange,
	minimum size =0.01cm,
	label={}] (R12) at (0.3+2,1.4-2.5){};

\node [draw,
	circle, fill = black,
	minimum size =0.01cm,
	label={}] (R13) at (0.3*3+2,1.4-2.5){};

\path [-] (Ad) edge[bend right=0, dashed, draw=blue,very thick] node { } (C2);
\path [->] (C2) edge[bend right=0, draw=red,very thick] node { } (Bf);
\path [->] (Bf) edge[bend right=0, draw=red,very thick] node { } (Dx);
\path [-] (Ax) edge[bend right=0, dashed, draw=blue,very thick] node { } (Rx);

\end{tikzpicture}
\caption{Urn model with recombination. An element (the red ball) that has not been drawn from the adjacent possible $\cal{U}$ is discovered by recombination of two of the available elements $\cal{S}$ (the black and orange balls) in a subspace that the agent focuses on. The possible recombinations of length two constitute the set $\cal{R}$. At time $t+1$ the red ball is added to the set of drawn elements $\cal{S}$. $\rho$ copies of the red ball and $\nu$ previously unavailable colors are added to the adjacent possible $\cal{U}$. In the illustration the agent searches among 4 colors and combinations of length 2.}
\label{fig:urnmodel}
\end{figure*}
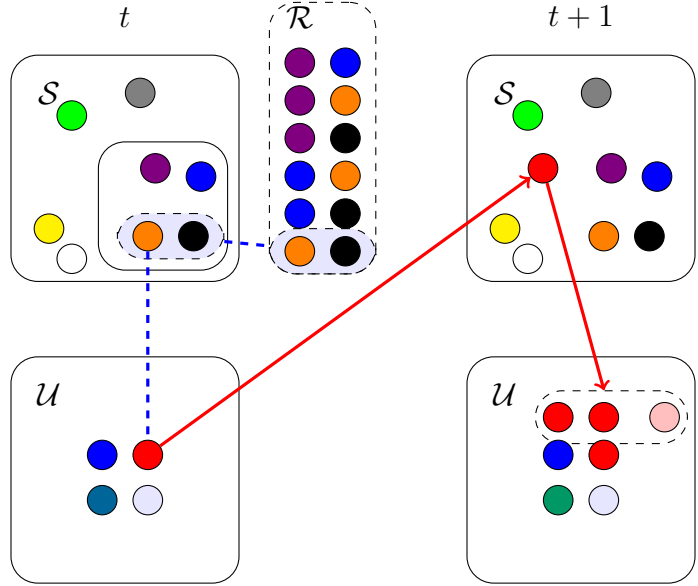

A view of innovation as recombinant naturally suggests that we should look at the set of all possible combinations, whose size is $\sum_k \alpha_L \frac{D!}{L! (D-L)!}$, where $\alpha_L$ is the relative frequency of an innovation among combinations of length $L$.   This formulation underlies e.g., \citet{weitzman1998} and \citet{koppl2021}. However, organizations face well-known constraints to search and innovation that motivate a more flexible and simple approach in formulating the adjacent possible. 

Firstly, facing costs of integration, search complexity should be limited to some maximal product length $\lambda$. If $\lambda$ is small relative to $D$  the sum is well approximated by only the largest product length $\frac{D!}{\lambda! (D-\lambda)!}$. 

Secondly, the organizations' ability to recognize, assimilate, and apply new knowledge inputs, viz. their absorptive capacity \citep{cohen1990}, will determine the scope of search activities (cf. \citealt{katila2002}).
Organizations may hence have a `window' of product types that it currently produces (for a similar argument, compare \citealt{vandam2020}.  Moreover, resource constraints or the ``ability of the research facility to test or to process the materials'' \citep[p.~353]{weitzman1998}  may imply that search takes place in a subset of possible knowledge recombinations. This effective set of recombinations is denoted $\cal{R}$. In other words, agents do not necessarily search for combinations among \emph{all} $D$ knowledge types in their portfolio, but rather among a set of currently searched elements $D^* \leq D$.  

With these modifications, the relevant space of adjacent possible product types can be written
\begin{equation}
\vert {\cal{U}}_D \vert = \nu  \frac{(D^* !)}{\lambda! (D^*-\lambda)!)}=\nu \vert {\cal{R}} \vert 
\end{equation} where $\lambda$ is the number of products that are recombined, and $\vert {\cal{R}}\vert$ is the number of recombinations of different knowledge types searched by an organization. $\nu$ is the fraction of recombinations that lead to the discovery of new elements. 

To complete the framework, we also need to consider the fact that organizations also rely on their experience in given fields to produce \emph{product improvements}, viz. multiple copies of a color. Both entirely new product types and product improvements may lead to product improvements in other fields. Every innovation made can be recombined with $D^*-1$ elements within the search space and has a probability of $D^* /D$ to be among the product types that are actively searched. Every innovation made then contributes with $\frac{D^*}{D} \frac{(D^*-1)!}{(\lambda-1)!(D^* - 1 -(\lambda-1))!} $ new possibilities. With $k$ cumulative innovations the total number of adjacent possible improvements becomes, with some simplification:
\begin{equation}
{\cal{U}_I} = \rho \frac{k}{D} \frac{D^*!}{(\lambda)!(D^* -\lambda)!} = \rho \frac{k}{D} \vert{\cal{R}}\vert
\end{equation} where $\rho$ expresses the fraction of recombinations that lead to new product improvements.

The total number of effectively adjacent possible innovations, viz. improvements and new product types, is of course given by

\begin{equation}
|{\cal{U}}| = |{\cal{U}_I}| + |{\cal{U}_D}| = \left( \nu + \rho \frac{k}{D} \right) \vert{\cal{R}}\vert
\end{equation}  

\subsection{Long-run rate of innovation}

A key tenet of the theory of the adjacent possible is that the rate of introduction of novelties,  here both new product types and product improvements, should be related to the size of the effective adjacent possible $\frac{dk}{dt}=|{\cal{U}}|$ \citep{koppl2021, cortes2022}.

\begin{figure*}
\centering
\includegraphics[scale =1.0]{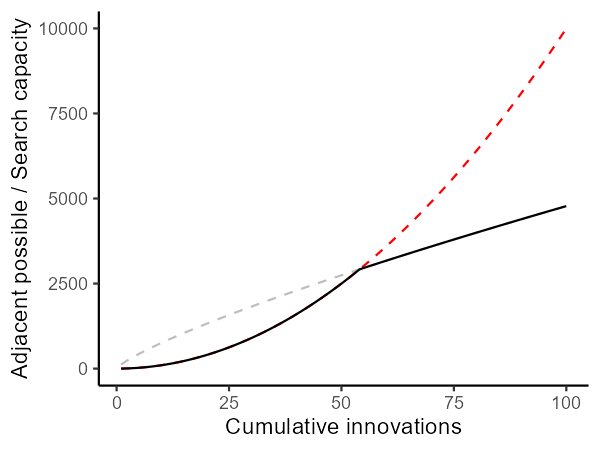}
\caption{Stylized illustration of how the rate of innovation (black line) is determined by the constraint of either the size of the adjacent possible (red dashed line) that grows in a super-linear fashion, or resource constraints (the grey dashed line) that is sublinear or linear in the cumulative number of innovations \citep{weitzman1998}. }
\label{fig:weakrecombination}
\end{figure*}

Following \citet{weitzman1998} there are, however, two extreme cases depending on resources available, illustrated in Figure \ref{fig:weakrecombination}. The rate of innovation is initially limited by the size of the adjacent possible, but then becomes constrained by resources available for search in the limit. In the \emph{strong version} of the argument, organizations have enough absorptive and search capacity to explore all, or a constant fraction of, the possible recombinations, such that $D^*=D$. In general, the rate of innovation can be re-expressed as a superlinear function of cumulative innovations by exploiting that $D$ must grow linearly or sub-linearly with $k$. Notably, under Heaps' law (see section \ref{sec:theoryheaps}):

\begin{equation}
\frac{dk}{dt} \sim k^{1+\frac{\nu}{\rho }(\lambda -1)}
\end{equation} suggesting that the rate of innovation, when unchecked, is superlinear in the cumulative number of innovations ($k$). In the most extreme case, the recombination length is unconstrained $\lambda = D$, leading to superexponential and explosive behavior (compare \cite*{cortes2022, koppl2021}), but even when $\lambda$ is a constant, this equation suggests ``winner-takes-all'' dynamics, where one firm dominate innovation activity in the limit.

In the other case, organizations can only make innovations as fast as they can secure a living from doing so (Kauffman, p. 156), and there are important resource constraints that make it difficult to search all possible combinations and costly to explore and integrate new types of knowledge. 

\begin{figure}
\centering
  \begin{tikzpicture}
    \draw[decoration={random steps, amplitude=1mm}, decorate] (2.5,0) circle (2);
    \draw[dashed, decoration={random steps, amplitude=1mm}, decorate] (3.5,0) circle (2);
      \node at (5,-3) {$D^*(t)$};
      \node at (1,-3) {$D^*(t+1)$};
      \node at (3,0) {$repeated$};
      \node at (1,0) {$new$};
  \end{tikzpicture}
  \caption{Two sets of elements searched by a firms at period $t$ and $t+1$. The new search space is defined by adding $new$ elements through exploration and repeating some elements from the earlier search space through exploitation.}
  \label{fig:search}
  \end{figure}
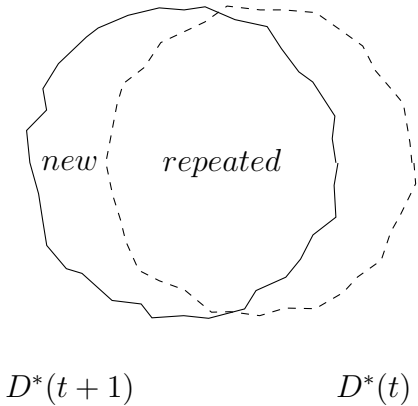

To derive an equation for this \emph{weak version} of recombination, we note that adjacent possible is, by definition, directly linked to search behavior, specifically to search scope and search depth \citep{katila2002}. This idea is illustrated in Fig. \ref{fig:search}. As before, firms search a set of elements. Exploitation means that organizations repeat elements they have searched before, and search depth is here defined as the fraction of elements that are repeated from earlier search $d = \frac{ rep }{ D^*}$ (compare \cite{katila2002}). Exploration means extending search to new elements and search scope is here defined as the fraction of new elements $s = \frac{new}{new+rep}$. It follows directly that one can express the search space in terms of scope and depth:

\begin{equation}
D^*(t+1) \equiv \left(\frac{d}{1-s} \right) D^* (t)
\end{equation} which can be expanded into

\begin{equation}
D^* = \prod_{j}^t \left( \frac{d_j }{1-s_j} \right) 
\end{equation} expressing the size of the search space as the cumulative impact of search scope and search depth. The equation is defined for $0 \leq d \leq 1$ and $0 \leq s < 1$.

If there are resource constraints, e.g., costs of integration of new knowledge, search scope and search depth will gravitate towards some long-run mean. Appendix \ref{sec:bianconibarabasi} shows that this produces a long run rate of innovation, which is equivalent to the Bianconi-Barabási model \citep{bianconi2001} with a \emph{linear} attachment kernel. The equation has a ``fitness parameter'' $\eta_i$ for each organization $i$ governed by the limiting rates of growth of search scope and search depth $\eta_i \propto \ln d_i - \ln (1-s_i)$. In full, 

\begin{equation}
\frac{dk}{dt} \propto \eta_i k
\label{eq:bb}
\end{equation}

The solution to this differential equation suggests that innovation within organizations follows exponential curves in the long run, rather than super-exponential patterns. This equation also suggests that organizations that are able to maintain high limiting rates of search scope and search depth will have high long-run rates of innovation.

This discussion can be summarized in the following hypotheses:

\emph{Hypothesis 1. The rate of innovation is initially super-linear (strong recombination), but eventually linear in the cumulative number of innovations (weak recombination).} 


\subsection{Distribution of innovation across organizations}

The distribution of innovations depends on the exponent of $k_i$ (see Appendix \ref{sec:discussion} for derivations). Sub-linear attachment kernels would yield a stretched exponential distribution. Linear attachment kernels yield a power law distribution with exponent $\approx -2$ (as derived in Appendix \ref{sec:distribution}), whereas a super-linear attachment kernel leads to winner-takes all distributions (see \citealp{krapivsky2001}). Since in the limit, we expect a linear attachment kernel, it follows that the overall distribution of innovations across organizations should follow a power-law with exponent $\approx -2$. However, this prediction is based on assuming that random elements across firms are negligible. As briefly discussed in Appendix \ref{sec:distribution}, varying growth rates across firms in the Bianconi-Barabási equation may also produce a log-normal distribution in the limit. 

\emph{Hypothesis 1b. The distribution of innovations across organizations follows a power law (with exponent approximately $-2$).}

\subsection{Product diversification}
\label{sec:theoryheaps}
The rate of product diversification should on average be given by the fraction of new product types (unique colors) in the adjacent possible:
\begin{equation}
\frac{dD}{dk} = \frac{\vert \cal{U}_D\vert}{\vert \cal{U} \vert} = \frac{\nu \vert {\cal{R}} \vert}{\nu \vert {\cal{R}}\vert + \rho \frac{k}{D} \vert {\cal{R}} \vert}
\end{equation}

Multiplying numerator and denominator by $D / \vert \cal{R} \vert$, we can rewrite in a simpler form and recover the dynamic equation from \cite{tria2014}:
\begin{equation}
\frac{dD}{dk} = \frac{\nu D}{\nu D + \rho k}
\label{eq:diffeq}
\end{equation}
This equation gives different solutions depending on parameter values $\nu$ and $\rho$ (see Appendix \ref{sec:heaps} for full derivation of expected dynamics). 

We are interested in the cases when $\nu < \rho$, what may be called a deepening search regime, and when $\nu > \rho$, what may be called widening search regime (paraphrasing \citealp{breschi2000}, on technological regimes). In the former case, it is more difficult (or costly) to discover entirely new types of products, and exploitation will dominate search activity. In the latter, it is more likely (or less costly) to discover new types of products, due to new opportunities or exhausted opportunities within a current technological trajectory. 

The received literature unanimously suggests that the former situation is the norm \citep{dosi1988, march1991}, but widening search patterns among organizations may emerge under episodes of strong external pressure and during paradigm shifts.
Since our interest lies in the long-run dynamics, one may expect a deepening search regime to dominate the empirical picture. In a deepening search regime, the differential equation \ref{eq:diffeq} produces Heaps' law \citep{heaps1978} with 

\begin{equation}
D \sim (\rho - \nu)^{\nu / \rho} k^{\nu/\rho}
\end{equation} For the widening search regime, product diversity approaches a fixed share of the cumulative number of innovations (see Appendix \ref{sec:heaps}). 

\emph{Hypothesis 2. In the long run, innovation activity generally takes place in a deepening search regime and the rate of product diversification follows Heaps' law.}

\subsection{Product space}

The theory of the adjacent possible suggests that the history of innovation determines the long-run rate of innovation and diversification. When it comes to more detailed predictions, Kauffman cautions that that the adjacent possible is ``unprestatable'', and, e.g., as regards the biosphere, ``we cannot mathematize [its] detailed becoming'' \citep[p. 3]{kauffman2014}. 
On the other hand, there are studies that propose that the space of diversification of products or skills is characterized by constraints. 
This type of dynamics has been explored previously, e.g., through constructing a ``product space'' based on skill-relatedness among firms \citep{neffke2013} and a global product space based on trade networks and the co-production of goods \citep{hidalgo2007, mealy2022}. 
These studies have shown that depending on where a firm or country is in the product space it is more or less easy to reach the core of the network, which e.g., explains the difficulties of poorer countries to develop sophisticated products \citep{hidalgo2007}.

In particular, if the notion of the adjacent possible is empirically relevant to organizational innovation, one should expect that the structure of cross-product relations lets us predict future diversifications of firms. In other words, following previous literature one may expect as a main hypothesis that the product space predicts future innovations and in particular what products firms diversify into. 

Following \citet{hidalgo2007}, one may formulate the hypothesis that the probability of organizations to diversify into a product field will depend on the proximity of an organization to that field given the history of products it has produced before:

\emph{Hypothesis 3. The structure of the product space predicts future innovations, both exploitation and exploration (new types of products).}

This will be explored by constructing a product space. For clarity of exposition and argument, the details of the calculations involved are introduced in the empirical section.

\section{METHODS AND DATA}
\label{sec:data}

\subsection{LBIO methodology}

To analyze long-run innovation dynamics across organizations, this study uses longitudinal data on new products and commercialized processes for the Swedish engineering industry for the period 1908-2016 (data after 1970 introduced in \citealt{sjoo2014} and \citealt{kander2019}, historical data introduced in \citealt{taalbi2021}). In recent years, work by historians and economists has made strides in producing long-run historical data enabling insights about innovation patterns and industrial dynamics, based on trade-journal literature, prize and awards data, machine learning methods applied to patent data or combinations thereof \citep{klepper2002, ortiz2018, taalbi2019, taalbi2021, kelly2021, capponi2022}. 

The data on innovation output used in the current paper is based on the screening of trade journals according to the so-called Literature Based Innovation Output (LBIO) method \citep{kleinknecht1993}. There are different views on what material and sections to include from the trade journals, but the currently employed innovation database rests firmly on the principle that material must be independent and edited. Consequently, the data does not include product announcements or advertisements, but edited articles typically written by journalists with some expertise in the field. In other words, advertisements and notification lists, based on press releases of firms, are not included.

The database, originally constructed for the period 1970-2007, is based on 15 trade journals covering the Swedish manufacturing industry and ICT services \citep{sjoo2014, kander2019}. The contemporary database includes almost 5,000 innovations, commercialized by Swedish firms, whose characteristics and innovation biographies are described in detail in the trade journal articles.

Assembly of this innovation data has advantages over patent and R and D statistics in capturing actual innovations, rather than inventions some of which are strategic or have little or no economic value. A drawback of the LBIO methodology is that manual collection of innovations is time and resource consuming. Another drawback is that in-house process innovations that do not enter into the market are underreported \citep{vanderpanne2007}.

To establish a long-run analysis of organizations' innovation activity, the present paper focuses on the two most important trade journals for period studied. \emph{Teknisk tidskrift} started in 1871, was published by the Swedish Association of Technologists, and was Sweden's foremost publisher of findings in engineering. In 1967, its weekly edition was continued under the name \emph{Ny Teknik}, published by the Swedish Association of Graduate Engineers. The second journal, \emph{Tidningen Verkstäderna} was founded in 1905 as the journal of the engineering industry's employer's association (Sveriges Verkstadsförening). Together these two journals reported 53\% of the total number of innovations during the period 1970-2016. 

It is important to note that the coverage of these journals has a decent overlap with other studies or lists of significant innovations. 54 out of 71 (76\%) major innovations in engineering listed by \citet{wallmark1991}, are included in the current dataset. Similarly, 40 out of 49 (81\%) engineering innovations listed in another list of major innovations \citep{sedig2002} are included in the current dataset. Those that have been found not to be covered in the current dataset are mostly innovations marketed by foreign companies and specialized machinery (e.g., for the paper and pulp, publishing and printing or chemical industries).

It is therefore reasonable to assume that the current dataset captures innovations by organizations active in the engineering industry and ICT services, except for some types of specialized suppliers of machinery. However, the product types that are covered in the two magazines are not limited to engineering products, but also include products across the board, including ceramics, wood and paper, chemicals and plastics and software. 

Nevertheless, diversification by engineering firms to these fields is underrepresented in the current data. Meanwhile, full data for the manufacturing and ICT services are available for the period 1970-2016. For this reason, as robustness checks, Appendix \ref{sec:extended} also presents the main results of this study using the full data for all 15 trade journals for the sub-period 1970-2016.

\subsection{Organizational boundaries and continuity}

\begin{figure*}[h]
\begin{subfigure}[b]{0.5\textwidth}
\caption{}
\label{fig:innovation_counts}
\includegraphics[scale =0.2]{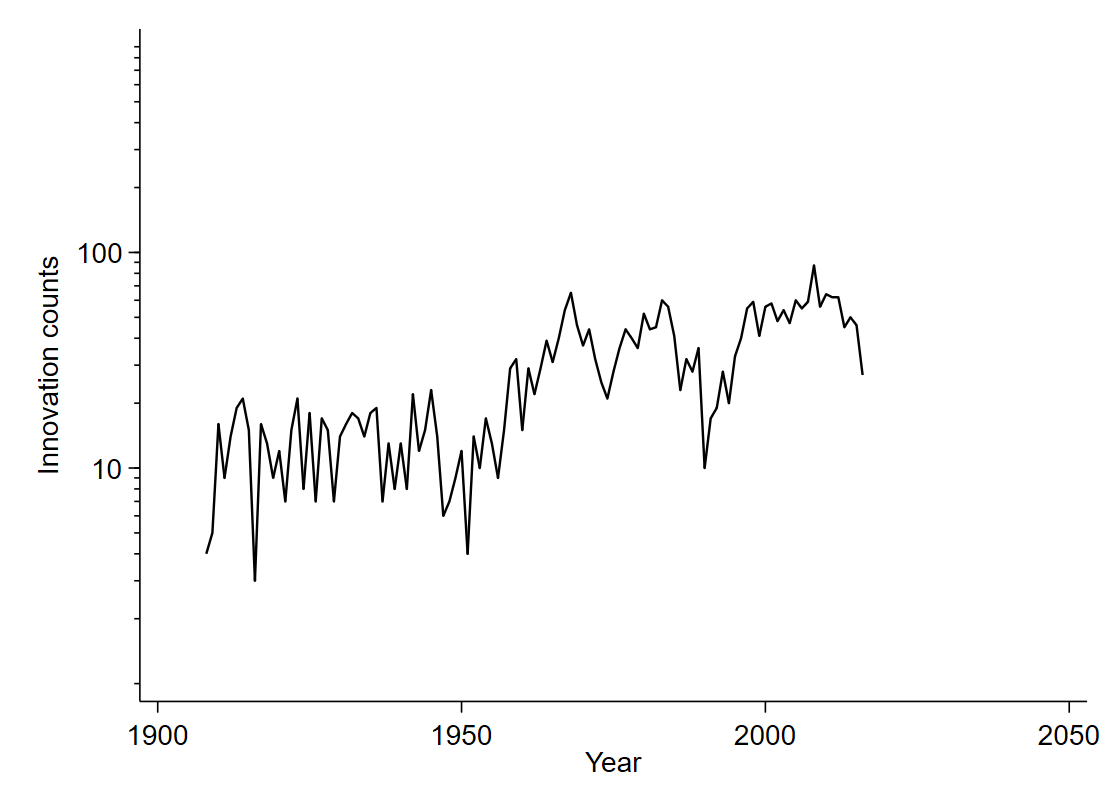}
\end{subfigure}
\begin{subfigure}[b]{0.5\textwidth}
\caption{}
\label{fig:entries}
\includegraphics[scale =0.2]{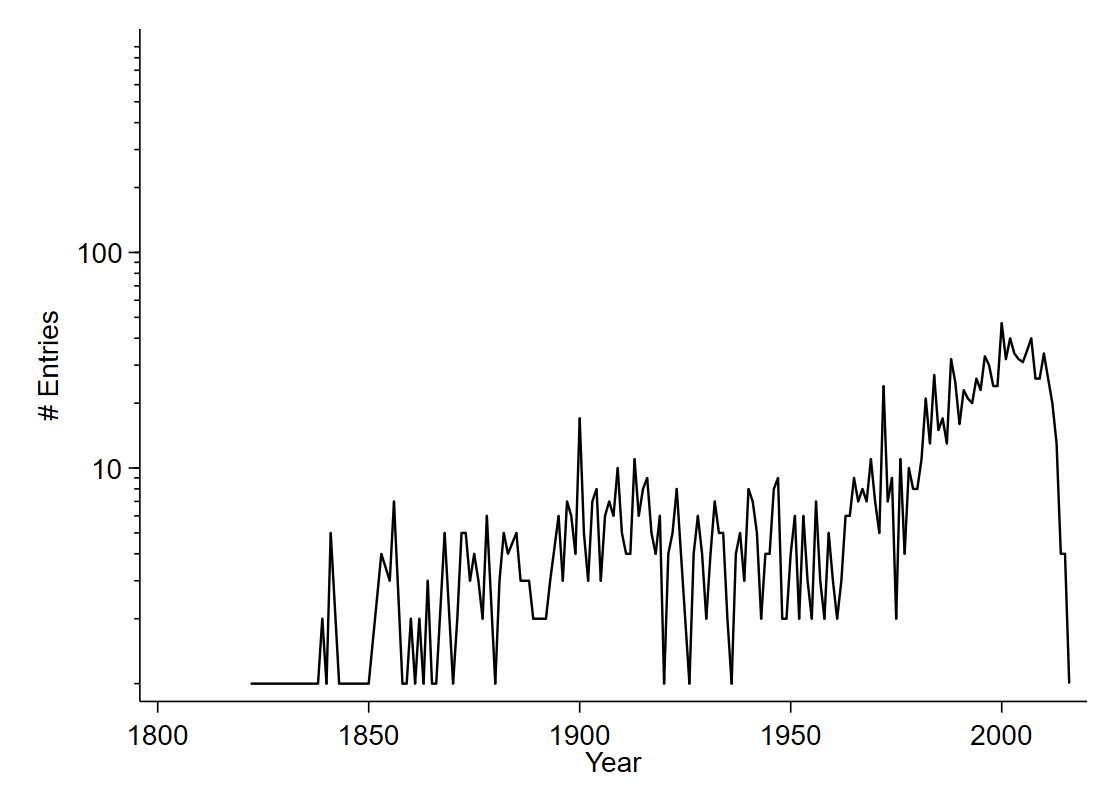}
\end{subfigure}
\caption{(a) Innovation counts (log) per year of commercialization, (b) Number of entries of innovative firms, by entry date. }
\end{figure*}

The data used in this paper concerns both information on firms and innovations in order to analyze the impact of the history of innovation on future innovation. Long-run series of firm-level innovation activity invariably encounters the problem of organizational change, viz. mergers, splits and acquisitions, why it is necessary to devise a definition of organizations and organizational continuity. Previous studies on Swedish firms have used flows of employees to trace mergers, acquisitions and firm survival \citep{eriksson2006, andersson2013}. Such data is however not available for the long time period studied here, wherefore this study employs a more heuristic approach based on company histories. Essentially, there are two pathways. One possibility is to define a firm as a single organizational unit that is discontinued under any merger, split or acquisition. This definition ensures that any organization refers to a coherent unit with coherent competencies, but has the downside of leading to biases and inconsistencies after the event. For example, a new merger combines competencies and capabilities from two pre-existing firms. Likewise, a split, does not render the new firms memoryless, and this strategy would underestimate historical experience of firms.

The other option is to collapse organizational units to a higher level when motivated by company histories. For example, Volvo is a corporate group consisting of several divisions with origins in diverse, originally independent, firms: the marine propulsion systems originate in AB Pentaverken, and the production of tractors originates in AB Bolinder-Munktell, in turn a merger of two previous firms. The main problem involved is that this creates a bias before merger events. This may, to some extent, be forgivable since firms rely on collaborations with similar firms or firms downstream in the supply chain, and such collaborations are frequently a predictor of later mergers and acquisitions.
In the absence of more refined aggregation methods, this paper uses the latter strategy.

\subsection{Variables used in this study}
\begin{table*}[htpb]
\caption{Summary statistics}
\footnotesize
\centering
\input{sumstat_2023.tex}
\label{tab:sumstat}
\end{table*}
\begin{table*}[htpb]
\caption{Correlation table}
\footnotesize
\input{corrstat_2023.tex}
\label{tab:correlations}
\end{table*}

The data used in this study is based on four variables: the product categories and commercialization year of innovations, and basic information on firm histories, specifically starting years and known exit dates. Based on the trade journal articles, each innovation is given a commercialization year. In the vast majority of cases, the trade journal article explicitly mentions a commercialization year explicitly. In the small minority of cases where the commercialization year was not mentioned, the year of the journal article was used as a proxy. Each innovation is also categorized into product groups as per ISIC Rev 3. 
 
The two journals together collect 3,086 innovations launched by 1,493 distinct organizations in the period 1908 - 2016. Most of these innovations were developed since the 1970s (Figure \ref{fig:innovation_counts}, see also \citealp{taalbi2021}). Similarly, most of these organizations started after the 1970s (Figure \ref{fig:entries}). For each of these innovations, a product code (ISIC Rev. 3) has been coded. The current work uses the 3-digit level codes to distinguish between product types. Data on firms' entry and exit dates were collected from Statistics Sweden's company registers for the period 1970-2016. All earlier data was collected from company histories (annual reports, firm biographies and Svensk Industrikalender). Since the data from Statistics Sweden does not capture splits and acquisitions, the data was cross-checked with these sources.

Finally, additional data has been collected for patented innovations \citep{taalbi2022} in order to control for the average growth of search depth and search scope, following \citet{katila2002}. This data covers slightly less than 50\% of the innovations for the period 1970-2016 \citep{taalbi2022}. These measures of search scope and search depth are defined on the basis of periods of five years. This study measures search scope $s_i$ as the number new 3-digit CPC classes cited by a firm in the period as a fraction of the total number of CPC classes cited by a firm in that period (cf. \citealt{katila2002}). The search depth $d_i$ is defined as the number of repeated CPC classes cited by a firm in the five year period as compared to the total number of CPC classes cited in the previous period. 

Following the theoretical framework, the geometric mean in search depth and search scope are calculated to test for a possible influence on the long-run rate of innovations. In order to keep to the theoretical motivation, $\ln \bar{d}$ was used as functional form for average search depth and $-\ln (1-\bar{s})$ for average search scope, although this renders regression coefficients slighlty more difficult to interpret.  Table \ref{tab:sumstat} gives summary statistics for the main regression variables, and Table \ref{tab:correlations} summarizes the Pearson correlation coefficients.

\section{RESULTS}
\label{sec:results}

\subsection{Main results}

\begin{table*}
\caption{Negative binomial regressions. Panels 1-4 are based on firm-year observations. Panels 4-6 are averages by the number of cumulative number of innovations.}
\footnotesize
\input{reg_2023.tex}
\label{tab:regressions}
\end{table*}

\begin{table*}
\caption{Baseline estimates of key parameters. Model 1 estimates the slope of the distribution of innovations. Model 2 estimates a linear baseline regression for the relative frequency of new product introductions across organizations. Model 3 estimates Heaps' law with product diversity (log) and cumulative innovations (log).	}
\label{tab:baseline}
\centering
\footnotesize
\input{simplewehc_2023.tex}
\end{table*}

\begin{table*}
\caption{Estimates of $\alpha$, goodness of fit (Kolmogorov-Smirnov statistic $KS$) for discrete power law distributions. The p-value tests the null hypothesis that the empirical distribution stems from a power law distribution. The log likelihood ratios $R$ and p-values test the null hypothesis that both a log-normal and power-law are equally far from the true distribution. Significant negative values imply that the log-normal has a better fit.}
\label{tab:ks}
\centering
\footnotesize
\centering
\begin{tabular}{rrrrrr}
  \hline
 & Power law & & & Log-normal & \\ 
Period & $\alpha$ & $KS$ & $p$ & $R$ & $p$ \\ 
  \hline
1970 & 2.55 & 0.01 & 1.00 & -2.86 & 0.00 \\ 
  1990 & 2.01 & 0.03 & 1.00 & -0.88 & 0.38 \\ 
  2010 & 2.07 & 0.04 & 0.99 & -0.65 & 0.52 \\ 
   \hline
\end{tabular}
\end{table*}

\begin{figure*}
\noindent\stackinset{l}{55pt}{b}{250pt}{\includegraphics[width=1in,trim=350 220 350 200,  clip]{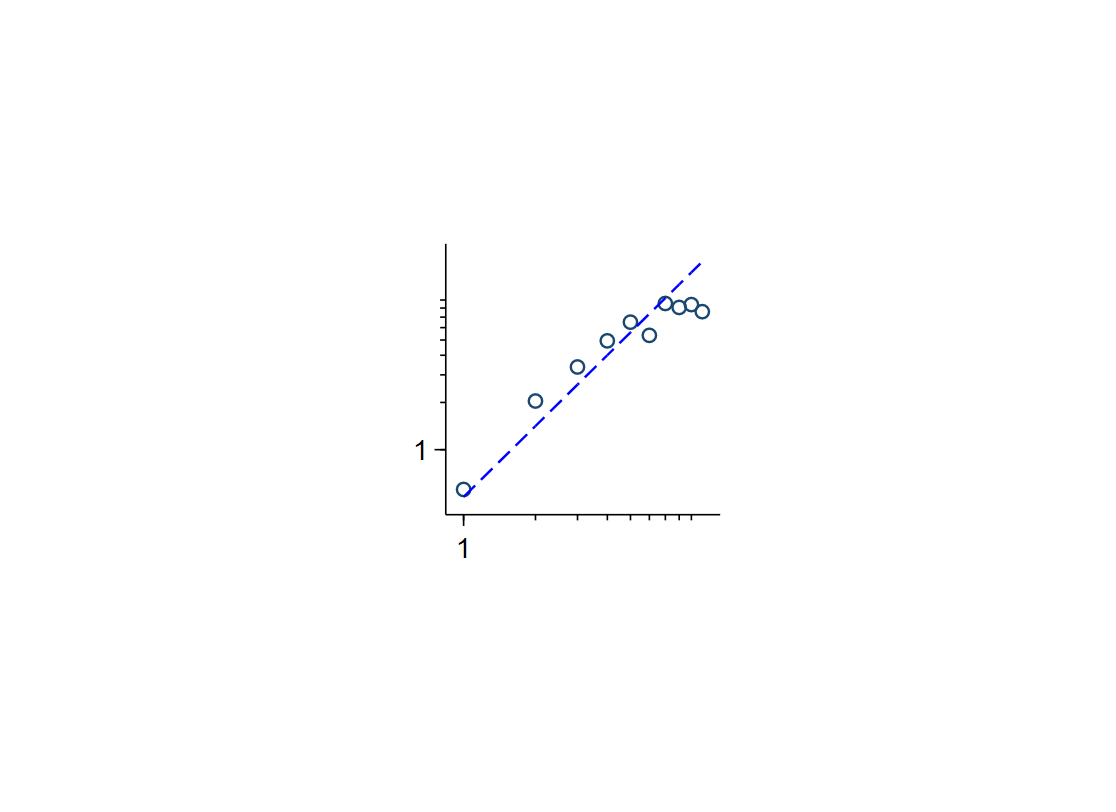}}
  {\includegraphics[width=\textwidth]{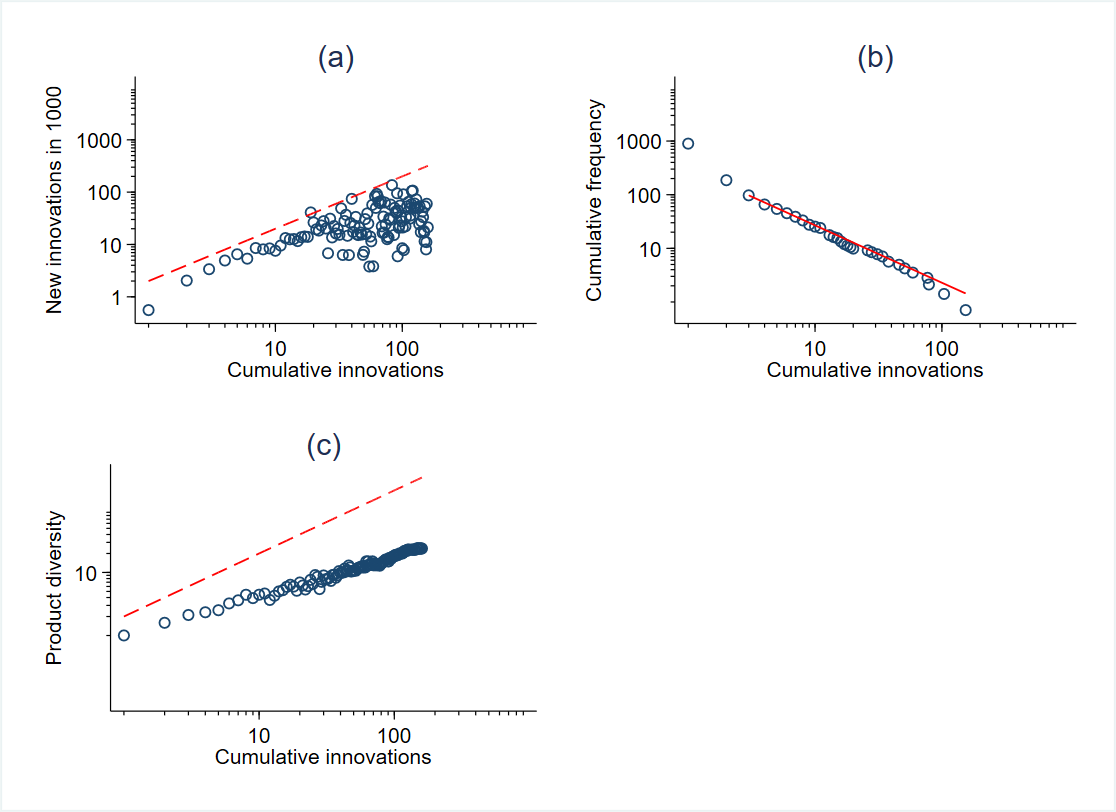}}
\centering
\caption{(a) Relative rate of new innovations by total cumulative innovations ($k$). Dashed lines demarcate linearity. The inset graph shows an early superlinear exponent of 1.5, (b) Distribution of the cumulative number of innovations $P(\kappa \geq k)$. Red line power law fit based on minimization of Kolmogorov-Statistics, (c) Product diversity ($D$) vs total cumulative innovation ($k$). Dashed line demarcates linearity.
}
\label{fig:panel}
\end{figure*}

The main results are presented in Figure \ref{fig:panel} and Tables -\ref{tab:regressions} - \ref{tab:baseline}. 

Figure \ref{fig:panel} first of all shows how new innovations scale with cumulative innovations. The rate of innovation is overall linear in cumulative innovations, but initially super-linear as the inset graph shows in Figure \ref{fig:panel}a. I also conduct formal tests of the the hypothesis that innovations are linear in cumulative innovations, and dependent on the average of search depth and scope. Since innovations are overdispersed count data, we use a standard negative binomial regression model, which models an independent variable $y$ as a function of a dependent variable $x$ as $Pr(Y = y \vert x) = \frac{\mu^{y} e^{-\mu}}{y!}
$ where $\mu = exp(\beta x + \epsilon )$ and $\epsilon$ being a Gamma-distributed random variable. If the independent variables are log transformed, $\beta$ describe log-log elasticities. Using equation \ref{eq:bb} we simply regress the number of innovations in (log) cumulative innovations, including controls for the size of the search space, and separately for average (log) search scope and average (log) search depth. For the search space size I used double-logs in keeping with \ref{eq:bb}. For the search depth and scope I used simple logarithms as approximations. Correlations suggest no multicollinearity (Table \ref{tab:correlations}).

The results, reported in Table \ref{tab:regressions} strongly suggest that innovation rates have a linear dependence on cumulative innovations. This applies in pooled negative binomial regressions (models 1-4) regardless of control variables. Models 5-7 collapse the data by the number of past cumulative innovations, focusing on the average patterns (compare \citealp{newman2001} and Figure \ref{fig:panel}a). Model 5 suggests a baseline sub-linear rate of innovation, but linearity is within the 95\% confidence interval when search depth and scope are included in Models 6-7. Taken together with Figure \ref{fig:panel}a the results strongly suggest that innovations are linear in cumulative innovations, as in the Bianconi-Barabási model.  
The results also generally corroborate the notion that the size of the search space impacts innovation rates, as the average the search space and average search depth and search scope have positive (and significant) effects on the rate of innovations. 

Table \ref{tab:baseline} estimates key parameters related to the distribution of innovations across organizations (model 1 and 2) and Heaps' law (model 3). Models 1 and 2 estimate the slope of the distribution of innovations across firms and the cumulative distribution of innovations respectively. The results agree with the expectation of a power law exponent of $2$ for the distribution across firms and $1$ for the corresponding cumulative distribution, shown in Figure \ref{fig:panel}b. 

As is well-known, however, these results do not necessarily confirm a power-law shape of the distribution, especially since econometric tests do not respect the requirement that the probability distribution must sum up to one. Following \citet{clauset2009}, Table \ref{tab:ks} tests the goodness of fit of the power-law distribution and compares with the alternative log-normal distribution. The results, for three benchmark distributions in 1970, 1990 and 2010 do not reject the null that the distribution of innovations across organizations stem from a power-law distribution. Follow-up tests using a log-normal cannot, however, exclude that the log-normal is an equally good fit, and perhaps better. Overall, these results are to a great extent in line with expectations from our theoretical framework, which propose either a power-law with exponent of $\approx 2$ or a log-normal distribution (see Appendix \ref{sec:distribution}).   

As discussed before, the data on products is based only on the engineering industry and the generic engineering trade journals. For this reason, as a robustness check, Appendix \ref{sec:extended} also presents the main results of this study using the full data for all 15 trade journals for the sub-period 1970-2016. Figure \ref{fig:panel_extended} and Table \ref{tab:regressions_extended} show that the results are very similar when taking to account product types not captured by the main data.

\subsection{Product space}

\label{data:productspace}

Finally, I explore the topology of the adjacent possible and the question of whether the product network of Swedish organizations can be used to predict in what product fields organizations develop future innovations, whether they repeat earlier product groups (exploitative innovation) or diversify to new ones (explorative innovation). The structure of ``cross-product'' relationships in the discovery of the adjacent possible, is here examined by constructing a product space of co-produced innovations within organizations, following earlier research \citep{hidalgo2007}. 

The product network can be constructed by mapping current innovations developed by an organization during a period of time to previous products produced, and estimating a proximity measure $\phi$. In the current analysis periods are defined on the basis of decades (1900s, 1910s, and so on). Let $k_{jl}$ be the cumulative number of times an organization $l$ has developed an innovation in product group $j$, and $k_l$ the cumulative number of innovations of the organization. The fraction $k_{jl}/k_{l}$ then represents the historical importance of product group $j$ for the organization in a period of time. The proximity of product $j$ to another product $i$ is obtained by taking a weighted sum over all firms producing product $i$, using the share of the firm's innovations in product group $i$ ($\Delta k_{il}/\Delta k_i$) as weights:
\begin{equation}
\phi_{ij}=\sum_l \frac{\Delta k_{il}}{\Delta k_i} \frac{k_{jl}}{k_l}
\end{equation}

For example, say firm $A$ has developed 80\% of all automotive innovations in a period and 50\% of its innovation portfolio consists of electric motor innovations. Firm B has developed 20\% of all automotive innovations in the period, but has no prior electric motor innovations. The proximity to automotive and electric motors is then $0.8 \times 0.5+0.2 \times 0$, viz. 0.4. 

Since this way of estimating proximity could in theory be driven by random processes, the networks only include estimates for which $\phi_{ij}$ is larger than the overall fraction of innovations in product $j$, viz. $p_j=k_j/k$. 
Even more specifically, firms should rely on recombinations of knowledge, in the sense that the more closely related products a firm has already made, the more likely they are to diversify into a given product. To test hypothesis 3, I use the product space to define the ``density'' (compare \citealp{hidalgo2007}) as
\begin{equation}
\omega_{jl} =\frac{ \sum_i k_{il} \phi_{ij}} {k_l}
\end{equation}
This weighted average expresses the extent to which an organization's product portfolio contains innovations in product groups that have a high proximity to a product $j$. 

\begin{table*}
\centering
\caption{Network statistics for sub-networks and the total period network 1908-2016.}
\footnotesize
\label{tab:networkstats}
\footnotesize{
\input{networkstats.tex}
}
\end{table*}

\begin{figure*}
\begin{subfigure}[b]{0.5\textwidth}
\caption{}
\label{fig:prediction_exploration}
\includegraphics[scale =0.2]{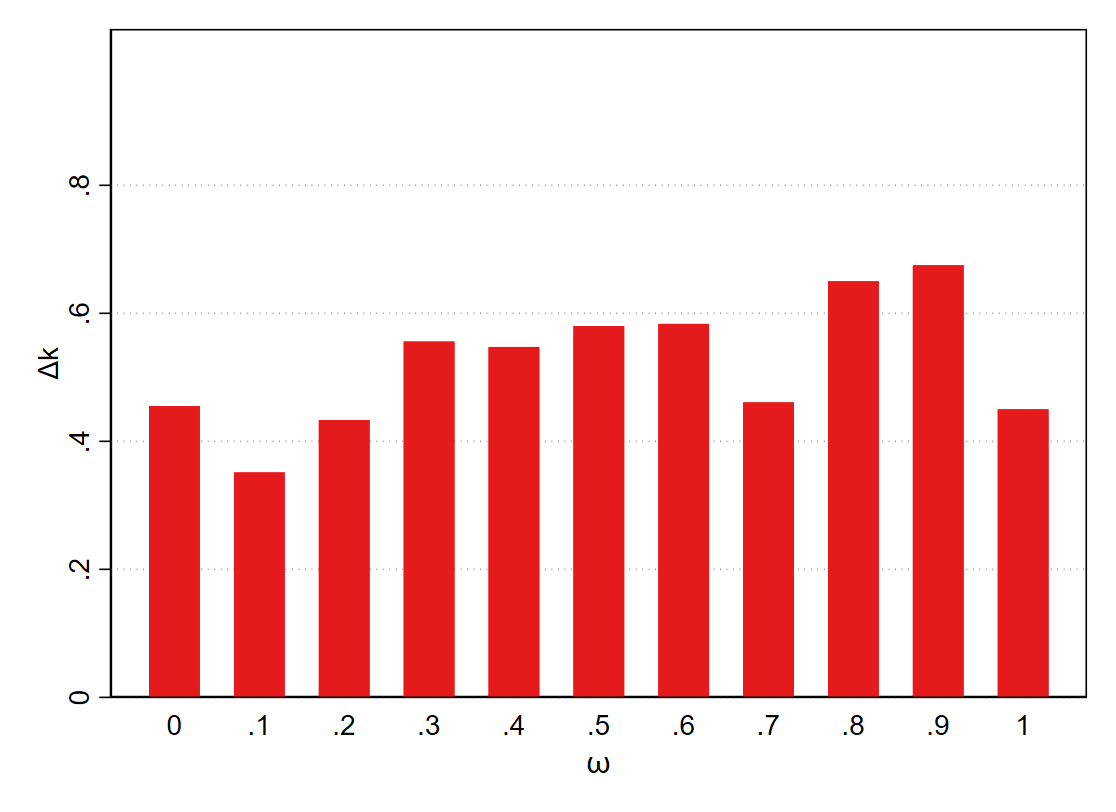}
\end{subfigure}
\begin{subfigure}[b]{0.5\textwidth}
\caption{}
\label{fig:prediction_exploitation}
\includegraphics[scale =0.2]{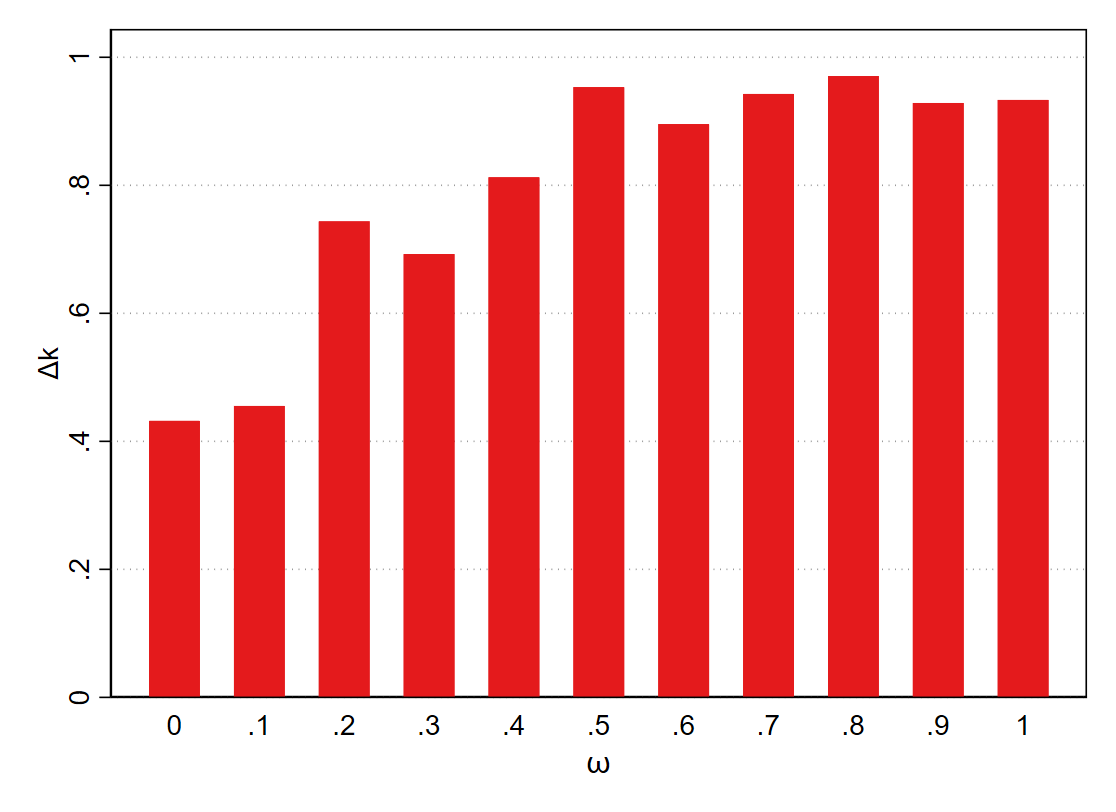}
\end{subfigure}
\caption{(a) Probability that an explorative innovations is in a given product group by the corresponding product density $\omega$ of a given firm, (b) probability that an exploitative innovations is in a given product group by the corresponding product density $\omega$ of a given firm.}
\label{fig:diversification}
\end{figure*}

\begin{figure*}
\centering
\includegraphics[scale=0.75]{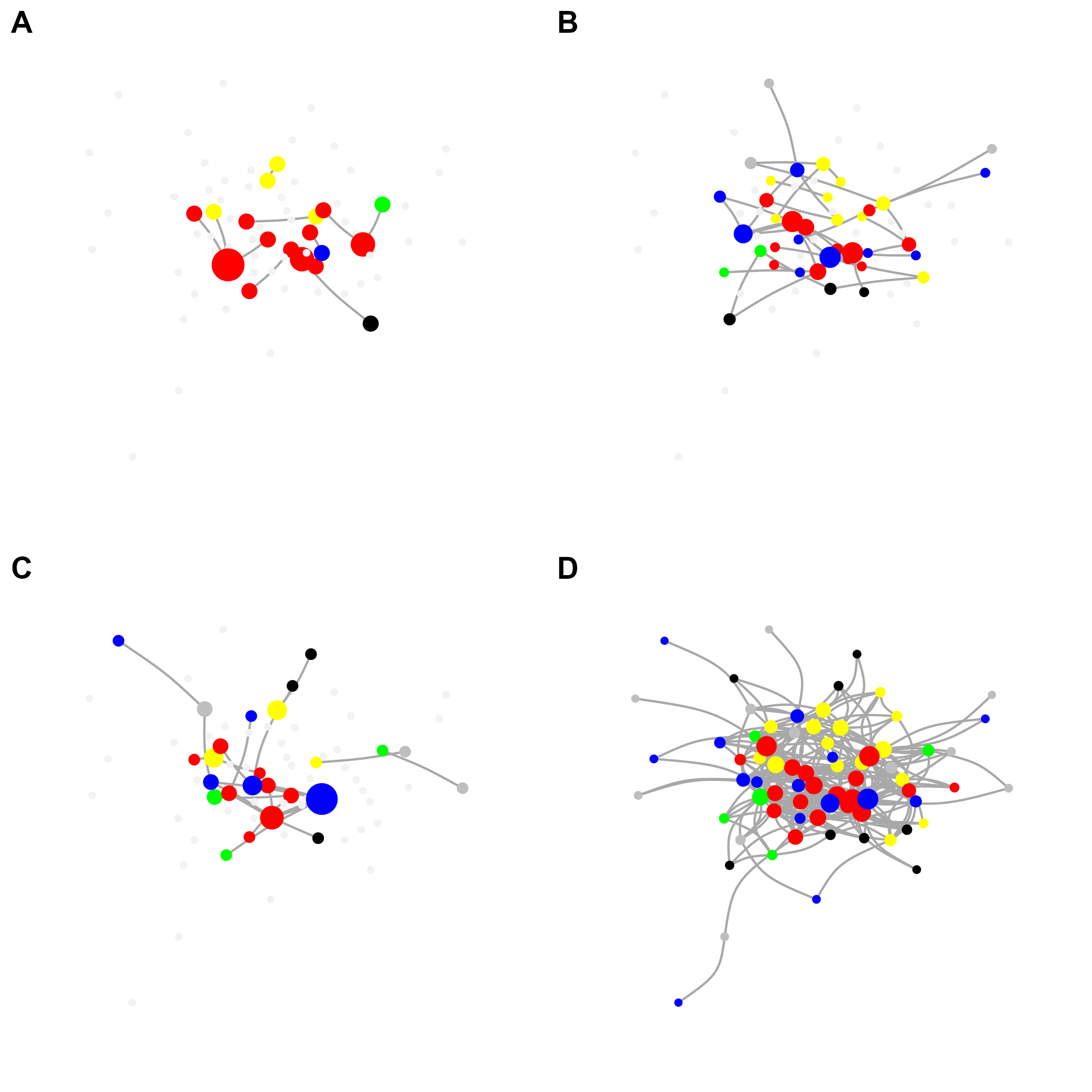}
\caption{Product networks for (a) 1930s, (b) 1970s, (c) 2010s, and (d) 1908-2016}
\label{fig:productnetworks}
\end{figure*}

\begin{figure*}
\begin{subfigure}[b]{0.5\textwidth}
\caption{}
\label{fig:predictions}
\includegraphics[scale =0.4]{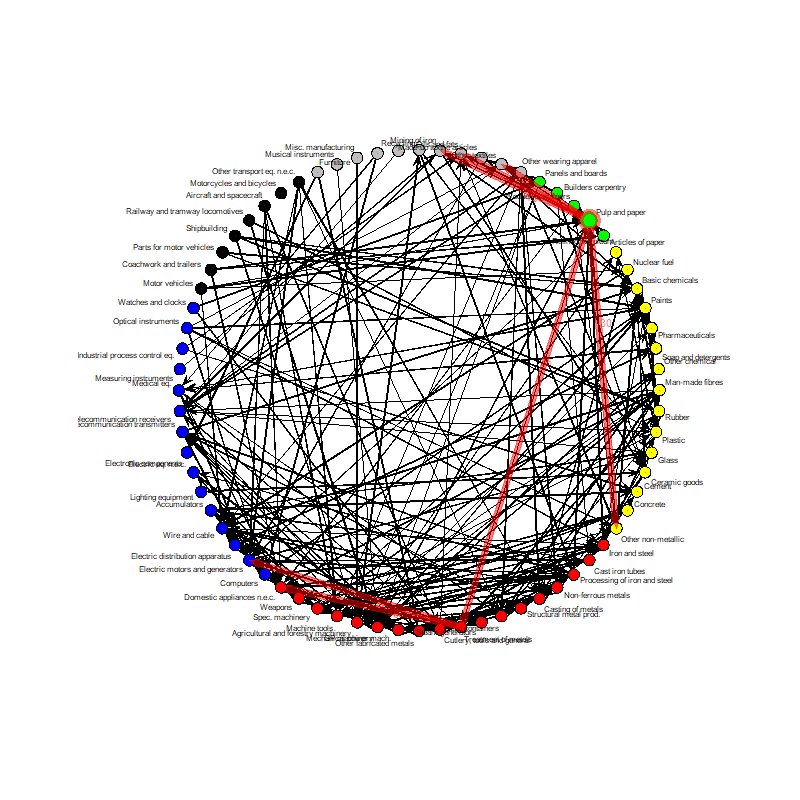}
\end{subfigure}
\begin{subfigure}[b]{0.5\textwidth}
\caption{}
\label{fig:predictions}
\includegraphics[scale =0.4]{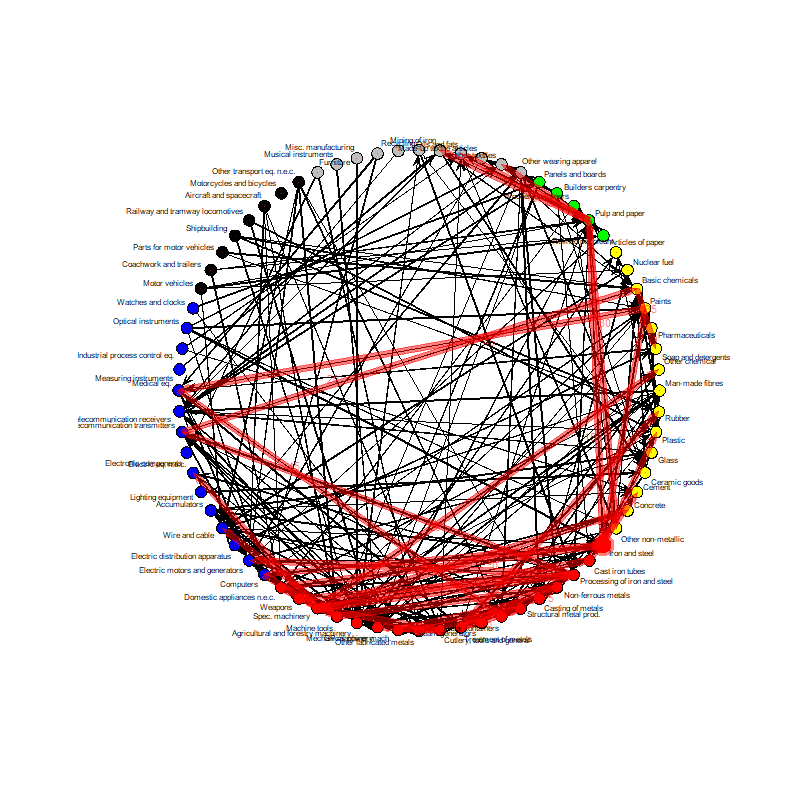}
\end{subfigure}
\begin{subfigure}[b]{0.5\textwidth}
\caption{}
\label{fig:predictions}
\includegraphics[scale =0.4]{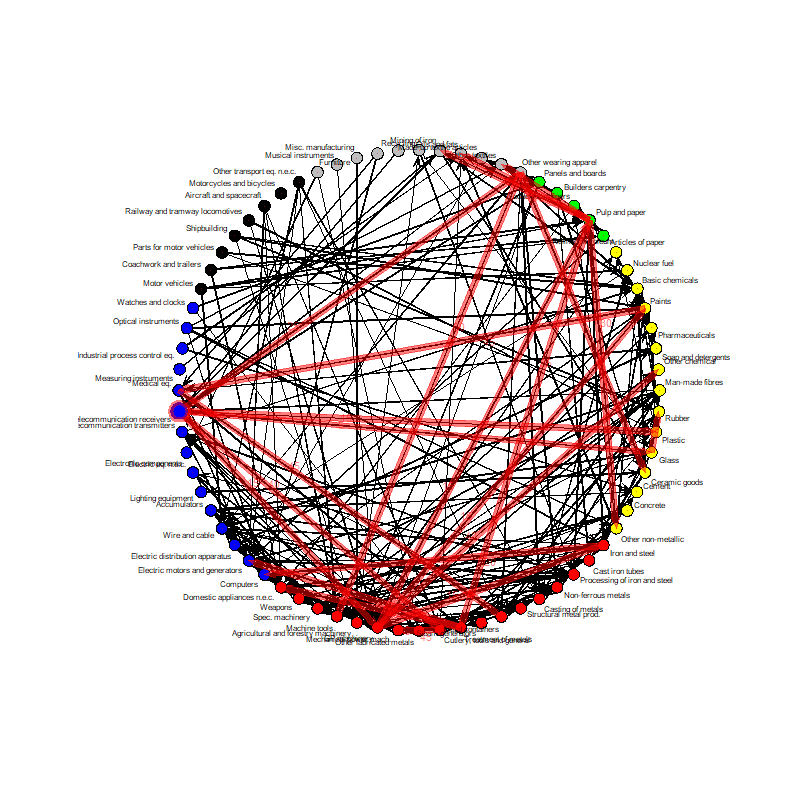}
\end{subfigure}
\begin{subfigure}[b]{0.5\textwidth}
\caption{}
\label{fig:predictions}
\includegraphics[scale =0.4]{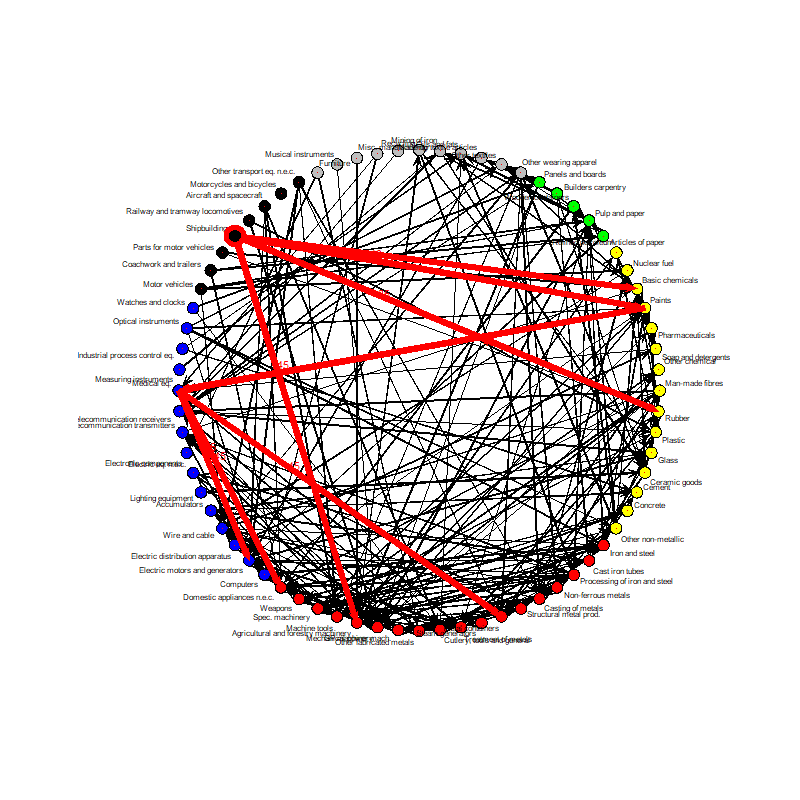}
\end{subfigure}
\caption{(a) Available direct and indirect pathways of product diversification 1970-2016 for firms initially active in (a) pulp and paper, (b) iron and steel, (c) radio, and (d) shipbuilding.}
\label{fig:pathways}
\end{figure*}

The main results are presented in Figures \ref{fig:productnetworks}-\ref{fig:pathways}. A first result is that the product density $\omega$, viz. the proximity of firms to a certain product field is, in general associated with the probability of a firm making an innovation in that field in the next period. The association is clearly positive, but slightly more patchy for entirely new diversifications, or ``explorative innovations''. This suggests that, while not precisely ``unprestatable'', the adjacent possible is not entirely governed by proximities in the product space.

It is also worth commenting on the structure of the product network. The network can be characterized as a relatively sparse small-world network (see Table \ref{tab:networkstats}). The small worldness statistic \citep{humphries2008}, comparing the transitivity and average path lengths to a random network, consequently suggests higher ``small worldness'' than a random network. This means that, although direct linkages are sparse, organizations in certain industries can relatively easily reach other product types through diversifying their product portfolio over time. Small-worldness and high reachability should be conducive to novelty generation \citep{bjorneborn2020}. 

However, there is substantial variation in the ability of certain industries to diversify, once link direction is taken into account. Figure \ref{fig:productnetworks} shows the product space for benchmark decades and the full period. The product space consists of a few central nodes, in the machinery (red) and ICT industries (blue), whereas other industries are more peripheral with less diversification paths. To illustrate this further, Figure \ref{fig:diversification} analyzes available pathways of diversification for producers in four industries that were severely affected by the oil crisis of the 1970s. Producers in the pulp and paper and shipbuilding industries had significantly less opportunities to diversify into new fields, than did producers of iron and steel products and radio producers.

\section{DISCUSSION}
\label{sec:discussion}

The results of this study suggest that the view of innovation as a process of recombination to discover the adjacent possible can be reconciled with empirical patterns. This is however only true in a weak form, where there are eventual constraints to innovation activity. The theoretical models predict, with reasonable accuracy, the empirical results as regards the distribution of innovations across organizations, the relative rate of new innovations and the rate of introduction of new types of products. The results also suggests that the structure of the product space can provide basic insights into how  organizations diversify their product portfolios over time.

This work shows at once the broad appeal and usefulness of the notion of innovation as a search for recombinations to find adjacent possible, but also important limitations. The arguably most surprising result of this study is that innovation may have a natural tendency to gravitate towards a linear dependence on cumulative innovations, and exponential growth curves. Here one must remind oneself that the analytical unit of this work is the single organization, often a private firm. These results do not hence exclude that super-linear patterns may appear in the macro-economy if, for example, the population of inventors or firms increases so as to make possible the discovery of untapped adjacent possible innovations beyond the capacity of individual incumbent firms. Ultimately, one may conjecture that checks to growth makes such scenarios intermittent rather than permanent, which again would engender exponential rather than super-exponential rates of innovation in the long run.

To better understand patterns and opportunities of diversification, the product space was also studied through mapping the co-production of innovations. Proximities in the product space displayed a general correlation to exploitative and explorative innovations, although weaker for explorative innovations (new product types). This suggests that the product space has important constraints, enabling some predictability. The product space also reveals a hierarchical pattern where complex product types, mainly machinery, transportation equipment and ICTs have high (out)degrees, and other products, like pulp and paper or pharmaceuticals, appear to imply high specialization and relatively low likelihood of diversification to other product groups (compare \citealp{hidalgo2007}). 
The results also suggest that the position of organizations in the product space matter for organizations' diversification paths and opportunities. Organizations with prior competence in machinery, transportation equipment or ICTs in their product portfolio could relatively easily explore other parts of the product space.

From a broader perspective, this study has explored the interaction between search behavior, industrial dynamics and the dynamics of organization' discovery of the adjacent possible and has suggested ways to theorize these connections. The framework is possible to extend to other situations where multiple agents explore the space of the adjacent possible through recombination, under behavioral or resource constraints. The results of this study suggest that this is a process that takes place in a kind of balance between the highly dynamic expansion of the adjacent possibilities and the checks and constraints that are placed on organizations and enforce exponential rates of innovation. These tensions are plausibly responsible for long-run outcomes across the life cycles of industries, where, early on, some organizations may cut ahead, building upon cumulated advances, but at the same time search, long-run resource constraints prevent the emergence of ``winner takes all'' situations, instead leaving enough space for new innovative entries. The dynamics of novelty creation and innovation would thus in this sense be fated to oscillate between cumulativity and renewal, continuity and disruption, and the rise and fall of major innovators.

\begin{singlespace}
\bibliographystyle{agsm}

\end{singlespace}

\section*{Acknowledgments}
I gratefully acknowledge funding support from Sweden's governmental agency for innovation systems, Vinnova (grant no 2014-06045 and 2020-01963).

\appendix
\counterwithin*{equation}{section}
\renewcommand\theequation{\thesection\arabic{equation}}

\section{Derivation of Heaps' law.}
\label{sec:heaps}
We start with the equation for the rate of introduction of new product types:
\begin{equation}
\frac{dD}{dk}=\frac{\nu D}{\nu D + \rho k}
\end{equation}

To solve this differential equation, we use substitution and plug in $z=D/k$, giving $\frac{dzk}{dk}=k dz/dk+z=\nu zk/(\nu zk+\rho k)$ and 
\begin{equation}
\int \frac{\nu z + \rho}{(\nu - \rho)z-\nu z^2} dz = \int \frac{1}{k} dk
\end{equation}
This gives 
$\rho/(\nu - \rho) \log z - \nu /(\nu - \rho) \log (\nu z - \nu + \rho) = \log k$
which can be rearranged to
\begin{equation}
\rho / \nu \log z - \log ( \nu z - \nu + \rho) = (\nu - \rho)/\nu \log k
\end{equation}

Solving out $D$,

\begin{equation}
\frac{\rho}{\nu} \log D - \log \left(\nu D/k - \nu + \rho \right) = \log k
\end{equation}
and rearranging
\begin{equation}
\rho / \nu \log D = \log \left(\nu D - \nu k \rho k \right)
\end{equation} which after taking the exponential and rearranging becomes
\begin{equation}
D^{\rho / \nu} - \nu D = (\rho - \nu) k
\label{eq:A1:deepening}
\end{equation} 
For large $D$ different results are obtained depending on the parameters. 
For deepening search regimes, $\nu < \rho$ , and one can ignore the second term on the left and derive
\begin{equation}
D \sim (\rho - \nu)^{\nu / \rho} k^{\nu / \rho}
\label{eq:A1:widening}
\end{equation} 
This scaling is sub-linear in $k$ and is known as Heaps' law. 
For widening search regimes, $\nu > \rho$ and one can ignore the first term on the left hand side and derive
\begin{equation}
D \sim \frac{(\nu - \rho)}{\nu} k
\end{equation} 
In other words, for deepening search regimes, the relative rate of new product types decreases over time. For widening search regimes, the relative rate of new product types approaches a fixed share of total new product introductions.

\section{Derivation of the Bianconi-Barabási equation}
\label{sec:bianconibarabasi}

The rate of innovation is given by the size of the effective adjacent possible, according to 

\begin{equation}
\frac{dk}{dt} = \left(\nu + \rho \frac{k}{D} \right) \vert {\cal{R}}\vert
\end{equation} with $\vert {\cal{R}}\vert = \frac{D^*!}{\lambda! (D-\lambda)!} \sim {D^*}^\lambda$

The search space  $D^*$ can be rewritten in terms of search scope $s$ and search depth $d$ as
\begin{equation}
D^* = \prod_t \frac{d_t}{1-s_t}
\end{equation}

We are looking for long-run dynamics and hence the equation can be written as  

\begin{equation}
\frac{dk}{dt} = \rho \frac{k}{D} \left( \prod_t \frac{d_t}{1-s_t} \right)^{\lambda}
\label{eq:A2:lrd}
\end{equation}

Using Heaps' law (equations \ref{eq:A1:deepening}-\ref{eq:A1:widening}) one can rewrite $D$ in general as limiting $D \sim \Lambda^{-1} k^{1-\gamma}$ as per equations $0<\gamma \leq 1$ . Hence $\frac{k}{D} = \Lambda k^{\gamma}$. 

Under constrained search, both $d$ and $s$ will vary within the interval $0 \leq d \leq 1$ and $0 \leq s < 1$. In the limit of large $t$ both scope and depth will approach long-run means $\bar{d}$ and $\bar{s}$. Using these two observations, one  can rewrite equation \ref{eq:A2:lrd} as 

\begin{equation}
\frac{dk}{dt} = \rho \Lambda k^{\gamma} \exp  \left(\eta t \right)
\label{eq:A2:dynamics}
\end{equation} where $\eta = \ln \left(\frac{\bar{d}}{1-\bar{s}} \right)$

If $\eta$ is zero this equation suggests that the rate of innovation is sub-linear in cumulative innovations. If $\eta$ is negative, viz. the organization decreases its search space in the limit, the rate of innovation will follow a sub-linear power law with an aging function, viz. a curvilinear pattern. If $\eta$ is positive, it is useful to note that this eq. \ref{eq:A2:dynamics} is equivalent to the Bianconi-Barabási model. From eq. \ref{eq:A2:dynamics} it is straightforward to use integration to get an expression for $k$ being equal to

\begin{equation}
 \begin{split}
k=\left(\rho \Lambda\right)^{\frac{1}{(1-\gamma)}} \left(\frac{\eta}{1-\gamma}\right)^{-\frac{1}{(1-\gamma)}}  \exp  \left(\frac{\eta}{1-\gamma} t \right)
\end{split}
\end{equation}

Plugging this back into equation \ref{eq:A2:dynamics}, or taking the derivative of the expression for k, gives the Bianconi-Barabási model:
 
\begin{equation}
\frac{dk}{dt} = \frac{\eta}{1-\gamma} k 
\label{eq:bianconi-barabasi}
\end{equation}

\section{Derivation of distribution of new product introductions}
\label{sec:distribution}

There are two approaches to derive the distribution of new product introductions across firms from the linear Bianconi-Barabási model discussed in Appendix \ref{sec:bianconibarabasi}. One is to simplify and ignore the fitness distributions, or assume that the deviations from a mean is small. In this case the framework leads to a power law distribution with exponent $-2$ (section A\ref{sec:distribution_2}). If, however, the effect of varying ``fitness'' is non-negligible, it can be shown that the cumulative number of innovations across fields will follow a log-normal distribution, as has been suggested in the discussion about Gibrat's law and firm growth (section A\ref{sec:distribution_1}). 

\subsection{Lognormal}
\label{sec:distribution_1}

The lognormal follows if one considers cross-sectional variation in $\eta$ and $\eta$ is i.i.d. across firms and over time with a given mean and standard deviation. In this case, the limiting distribution follows trivially from equation \ref{eq:bianconi-barabasi}. Simplifying slightly, we have

\begin{equation}
\frac{dk}{dt} \frac{1}{k} = \eta
\end{equation}

One may note that $\frac{d (\log k) }{dt} = \frac{dk}{dt} \frac{1}{k} $, which gives the equation

\begin{equation}
\log k = \int \eta dt
\end{equation}

The Central Limit Theorem of probability theory states that the sum of i.i.d variables approaches a normal distribution. $k$ then clearly will approach a lognormal distribution.

\subsection{Power law}
\label{sec:distribution_2}
Say we have $N$ independent organizations at time $t$. New entrants occur at a rate $m$ at each point in time. In the simplest model, with a linear or sub-linear attachment kernel, the probability of making a new innovation is
\begin{equation}
\Pi_k=\frac{k^\lambda}{\sum_k k^\lambda N_k }
\end{equation}

The distribution is given by first constructing the master equation for the number of firms with innovations $k$:
\begin{equation}
N_k (t+1) = N_k + \Pi_{k-1} N_{k-1} - \Pi_k N_k
\end{equation}

We define the number of firms as $N=mt +N_0$. The share of firms with $k$ innnovations is $P_k=N_k/N$. This means that
\begin{equation}
\frac{\partial N_k}{\partial t} = \frac{\partial N_k}{\partial N} \frac{\partial N}{\partial t} = m P_k
\end{equation}
which combined with the master equation gives
\begin{equation}
m P_k= N \Pi_{k-1} P_{k-1} - N \Pi_k P_k
\end{equation}
which can be rearranged to
\begin{equation}
P_k = \frac{N \Pi_{k-1}}{m + N \Pi_{k}} P_{k-1}
\end{equation}

In our case, we plug in attachment kernel and use $\mu = \sum_k \left( k^\lambda N_k\right) / N$ to simplify to:
\begin{equation}
P_k = \mu^{-1} \frac{\left( k-1 \right)^{\lambda}}{m + \mu^{-1} k^\lambda} P_{k-1}
\end{equation}

Or
\begin{equation}
P_k \frac{1}{k^\lambda} \prod_j \frac{j^\lambda}{\mu m + j^\lambda} P_0
\end{equation}

We now have two cases. 
If $\lambda =1$  

\begin{equation}
P_k = \frac{1}{k} \frac{k!}{\left(\mu m + k\right)! } \mu m ! P_0
\end{equation}

Using the fact that the continuous approximations of $x! / (a+x)! \sim x^{-a}$,
this is similar to a power law

\begin{equation}
P_k \sim k^{-1-\mu m}
\end{equation} 

With $\lambda =1 $ and $\mu =\sum_k \left( k^\lambda N_k \right)/N = \frac{t}{mt} = \frac{1}{m}$ and hence 
\begin{equation}
P_k \sim k^{-2}
\end{equation} 

For sub-linear attachment kernels, i.e. $\lambda <1$, we can write the distribution as

\begin{equation}
\begin{split}
P_k = \frac{1}{k^\lambda } P_0 \exp \left( \sum_j \ln (\frac{j^\lambda }{\mu m + j^\lambda}\right) = \\ \frac{1}{k^\lambda} P_0 \exp \sum_j \left(- \ln (\mu m j^{-\lambda} +1) \right)
\end{split}
\end{equation}
If we take the series representation \begin{equation}
\ln \left( \mu m j^{-\lambda}+1\right) =  - \sum_j \frac{\left(-1^{i} (\mu mj^{-\lambda})^i  \right)} {i} 
\end{equation}
and noting that the sum vanishes for large $i$ we can take the integral and obtain a stretched exponential:
\begin{equation}
\begin{split}
P_k = \frac{1}{k^\lambda} P_0 \exp \int -\left(\mu m j^{-\lambda} \right) dj = \\ \frac{1}{k^\lambda} P_0 \exp \left(\frac{-1}{1-\lambda} \mu m k^{1-\lambda} \right)
\end{split}
\end{equation}

\onecolumn
\clearpage
\section{Results with extended data 1970-2016}
\label{sec:extended}

\begin{figure*}[htbp]
\noindent\stackinset{l}{55pt}{b}{250pt}{\includegraphics[width=1in,trim=350 220 350 200, 
 clip]{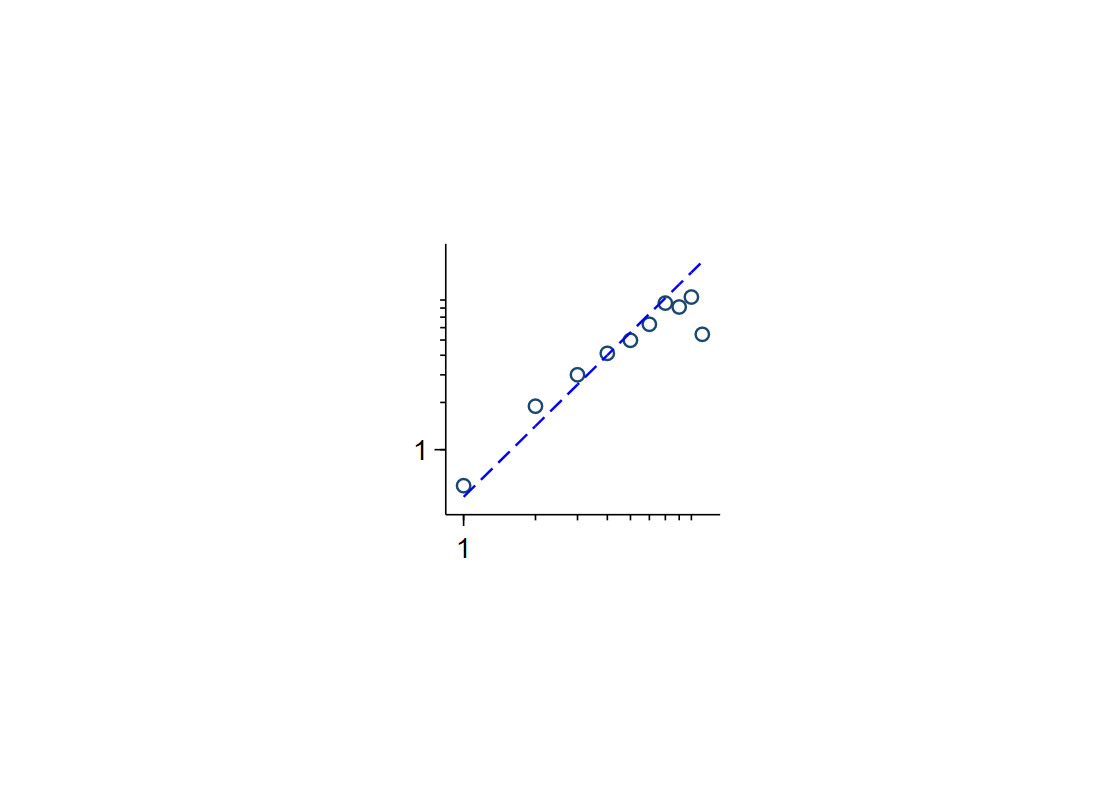}}
  {\includegraphics[width=\textwidth]{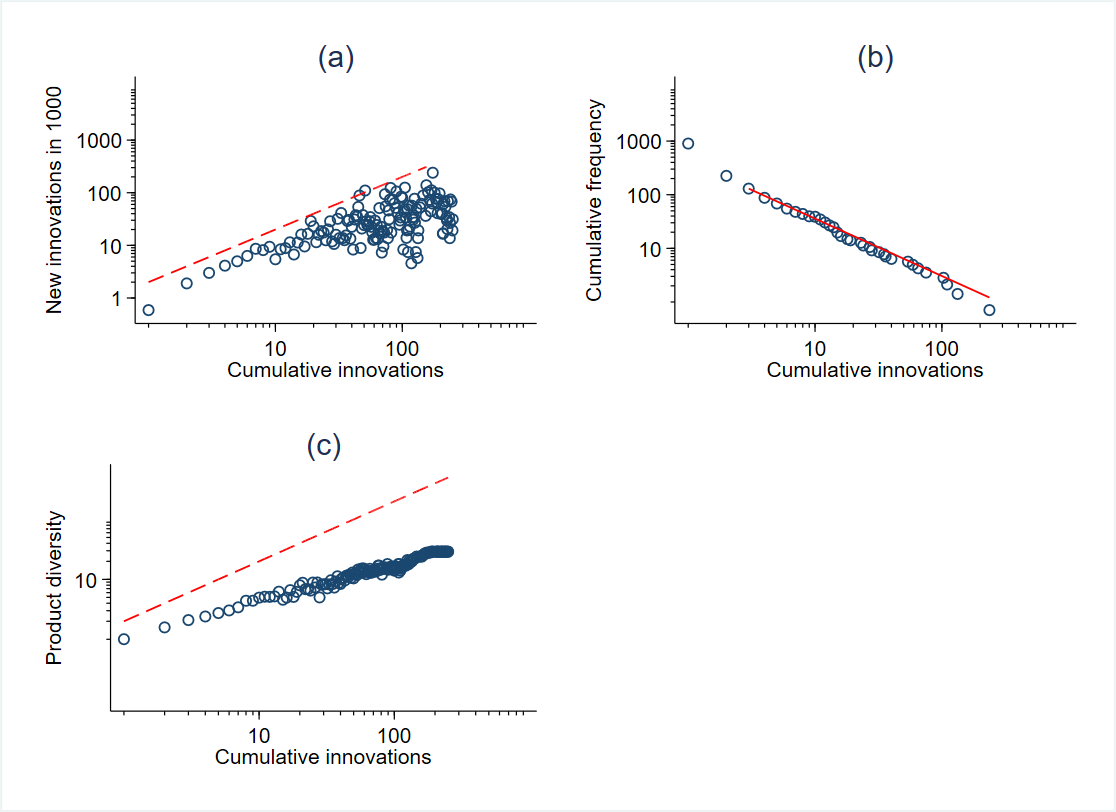}}
\centering
\caption{(a) Relative rate of new innovations by total cumulative innovations ($k$). Dashed lines demarcate linearity. The inset graph shows an early superlinear exponent of 1.5, (b) Distribution of the cumulative number of innovations $P(\kappa \geq k)$. Red line power law fit based on minimization of Kolmogorov-Statistics, (c) Product diversity ($D$) vs total cumulative innovation ($k$). Dashed line demarcates linearity.
}
\label{fig:panel_extended}
\end{figure*}

\begin{table*}[htbp]
\caption{Negative binomial regressions. Panels 1-4 are based on firm-year observations. Panels 4-6 are averages by the number of cumulative number of innovations.}
\footnotesize
\input{reg_2023_extra.tex}
\label{tab:regressions_extended}
\end{table*}

\end{document}

%% file: sumstat_2023.tex
\begin{tabular}{l c c c c c}\hline\hline
\multicolumn{1}{c}{\textbf{Variable}} & \textbf{Mean}
 & \textbf{Std. Dev.}& \textbf{Min.} &  \textbf{Max.} & \textbf{N}\\ \hline
New inno & 0.062 & 0.291 & 0 & 7 & 49897\\
Cumulative innovations (log) & 0.46 & 0.871 & 0 & 5.075 & 25843\\
Search depth (av.) & -0.187 & 0.622 & -5.744 & 0 & 11515\\
Search scope (av.) & 5.298 & 1.165 & 0.156 & 7.864 & 11515\\
Age & 42.565 & 57.063 & 0 & 436 & 49897\\
\hline\end{tabular}

%% file: corrstat_2023.tex
\begin{tabular}{l  c  c  c  c  c }\hline\hline
\multicolumn{1}{c}{Variables} &New inno&Cumulative inno. (log)&Search depth (av.)&Search scope (av.)&Age\\ \hline
New inno&1.000\\
Cumulative inno. (log)&0.360&1.000\\
Search depth (av.)&-0.144&-0.451&1.000\\
Search scope (av.)&-0.038&-0.244&0.327&1.000\\
Age&0.026&0.362&-0.209&-0.180&1.000\\
\hline \hline 
 \end{tabular}

%% file: reg_2023.tex
\begin{tabular}{lccccccc} \hline
 & (1) & (2) & (3) & (4) & (5) & (6) & (7) \\
VARIABLES & New inno & New inno & New inno & New inno & New inno & New inno & New inno \\ \hline
 &  &  &  &  &  &  &  \\
Cumulative innov. (log) & 0.997 & 0.994 & 0.970 & 0.997 & 0.645 & 0.856 & 1.024 \\
 & (0.0216) & (0.0320) & (0.0338) & (0.0220) & (0.123) & (0.190) & (0.203) \\
 & [0.000] & [0.000] & [0.000] & [0.000] & [0.000] & [0.000] & [0.000] \\
Search space (log) &  & 0.238 &  &  &  &  &  \\
 &  & (0.0627) &  &  &  &  &  \\
 &  & [0.000] &  &  &  &  &  \\
Search depth (av.) &  &  & 0.0561 &  &  & 0.0613 & 0.0753 \\
 &  &  & (0.0535) &  &  & (0.114) & (0.112) \\
 &  &  & [0.295] &  &  & [0.591] & [0.503] \\
Search scope (av.) &  &  & 0.0919 &  &  & 0.176 & 0.274 \\
 &  &  & (0.0382) &  &  & (0.0911) & (0.0948) \\
 &  &  & [0.016] &  &  & [0.054] & [0.004] \\
Age &  &  &  & -0.000556 &  &  & -0.0142 \\
 &  &  &  & (0.000551) &  &  & (0.00403) \\
 &  &  &  & [0.312] &  &  & [0.000] \\
Constant & -3.940 & -4.418 & -4.497 & -0.684 & -2.506 & -4.016 & -3.751 \\
 & (0.0456) & (0.135) & (0.224) & (0.848) & (0.542) & (0.910) & (0.949) \\
 & [0.000] & [0.000] & [0.000] & [0.420] & [0.000] & [0.000] & [0.000] \\
 &  &  &  &  &  &  &  \\
Observations & 25,843 & 9,469 & 9,506 & 25,843 & 126 & 125 & 125 \\
 Year FE & No & No & No & Yes & No & No & No \\ \hline
\multicolumn{8}{c}{ Standard errors in brackets} \\
\end{tabular}

%% file: simplewehc_2023.tex
\begin{tabular}{lccc} \hline
 & (1) & (2) & (3) \\
VARIABLES & $ P(k) $ (log) & CDF (log) & Product diversity (log) \\ \hline
 &  &  &  \\
Cumulative inno. (log) & -2.213 & -1.111 & 0.587 \\
 & (0.122) & (0.0323) & (0.00172) \\
 & [0.000] & [0.000] &  [0.000]\\
Constant & 6.453 & 5.953 & 0.00564 \\
 & (0.275) & (0.0960) & (0.00170) \\
 & [0.000] & [0.000] & [0.001] \\
 &  &  &  \\
Observations & 20 & 32 & 25,677 \\
 R-squared & 0.948 & 0.975 & 0.820 \\ \hline
\multicolumn{4}{c}{ Standard errors in brackets} \\
\end{tabular}

%% file: networkstats.tex
\centering
\begin{tabular}{rlrrrr}
  \hline
 & 1930 & 1970 & 2010 & 1908-2016 \\ 
  \hline
 N & 18.00 & 38.00 & 25.00 & 72.00 \\ 
   E & 11.00 & 41.00 & 23.00 & 335.00 \\ 
   Density (\%) & 3.59 & 2.92 & 3.83 & 6.55 \\ 
   Av. degree & 0.61 & 1.08 & 0.92 & 4.65 \\ 
   Std. degree & 0.85 & 1.50 & 1.35 & 4.57 \\ 
   Transitivity (\%) & 0.00 & 14.63 & 20.69 & 25.72 \\ 
   Av. path length & 1.08 & 2.34 & 1.24 & 2.81 \\ 
   Connectivity & 7.00 & 5.00 & 5.00 & 1.00 \\ 
   Biggest component & 4.00 & 22.00 & 13.00 & 72.00 \\ 
   Small worldness & 1.43 & 30.05 & 1.41 & 1.38 \\ 
   \hline
\end{tabular}

%% file: reg_2023_extra.tex
\begin{tabular}{lccccccc} \hline
 & (1) & (2) & (3) & (4) & (5) & (6) & (7) \\
VARIABLES & New inno & New inno & New inno & New inno & New inno & New inno & New inno \\ \hline
 &  &  &  &  &  &  &  \\
Cumulative innovations (log) & 0.970 & 1.007 & 0.979 & 0.989 & 0.697 & 0.979 & 1.145 \\
 & (0.0179) & (0.0263) & (0.0275) & (0.0191) & (0.0874) & (0.145) & (0.157) \\
 & [0.000] & [0.000] & [0.000] & [0.000] & [0.000] & [0.000] & [0.000] \\
Search space &  & 0.225 &  &  &  &  &  \\
 &  & (0.0495) &  &  &  &  &  \\
 &  & [0.000] &  &  &  &  &  \\
Search depth (av.) &  &  & 0.0359 &  &  & 0.127 & 0.238 \\
 &  &  & (0.0399) &  &  & (0.0939) & (0.0989) \\
 &  &  & [0.368] &  &  & [0.176] & [0.016] \\
Search scope (av.) &  &  & 0.100 &  &  & 0.196 & 0.232 \\
 &  &  & (0.0304) &  &  & (0.0680) & (0.0665) \\
 &  &  & [0.001] &  &  & [0.004] & [0.000] \\
Age &  &  &  & -0.000825 &  &  & -0.00913 \\
 &  &  &  & (0.000488) &  &  & (0.00293) \\
 &  &  &  & [0.091] &  &  & [0.002] \\
Constant & -3.725 & -4.166 & -4.306 & -0.679 & -2.492 & -4.317 & -4.016 \\
 & (0.0404) & (0.112) & (0.182) & (0.837) & (0.411) & (0.731) & (0.738) \\
 & [0.000] & [0.000] & [0.000] & [0.417] & [0.000] & [0.000] & [0.000] \\
 &  &  &  &  &  &  &  \\
Observations & 26,359 & 9,600 & 9,637 & 26,359 & 152 & 152 & 152 \\
 Year FE & No & No & No & Yes & No & No & No \\ \hline
\multicolumn{8}{c}{ Standard errors in brackets} \\
\end{tabular}